\documentclass[12pt,epsfig,epsf]{article}
\usepackage{amsmath}
\usepackage{cite}
\usepackage{slashed}
\usepackage{epsf,color,colordvi}
\usepackage{graphicx}
\usepackage{amsmath}
\usepackage{amssymb}
\usepackage{enumerate}
\usepackage{epsfig}
\usepackage{cite}
\usepackage{fontenc}
\usepackage{float}
\usepackage{xcolor}
\usepackage{booktabs}
\usepackage{tabularx}
\usepackage{multirow}
\usepackage{longtable}
\usepackage{lscape}
\usepackage{dcolumn}
\usepackage{rotating}
\usepackage[english]{babel}
\usepackage[autostyle]{csquotes}
\usepackage{hyperref}
\usepackage[normalem]{ulem}
\usepackage{color}
\usepackage{caption}
\usepackage{accents}

\newcommand {\ignore}[1]{}
\definecolor{darkgreen}{cmyk}{1,0,1,0.4}
\definecolor{brown}{cmyk}{0,0.8,1,0.2}
\definecolor{darkred}{cmyk}{0,1,1,0.2}

\def\nue{{\nu_e}}

\def\numu{{\nu_{\mu}}}
\def\numubar{{\bar{\nu}_{\mu}}}
\def\nutau{{\nu_{\tau}}}

\newcommand{\eg}{{\it e.g.}}
\newcommand{\ie}{{\it i.e.}}
\newcommand{\etc}{{\it etc.}}

\newcommand{\beq}{\begin{equation}}
\newcommand{\eeq}{\end{equation}}
\newcommand{\beqa}{\begin{eqnarray}}
\newcommand{\eeqa}{\end{eqnarray}}

\newcommand{\ta}{\theta_{12}}
\newcommand{\tb}{\theta_{13}}
\newcommand{\tc}{\theta_{23}}
\newcommand{\da}{\delta_{13}}
\newcommand{\ldm}{\Delta m_{31}^2}
\newcommand{\sdm}{\Delta m_{21}^2}
\newcommand{\rad}{R_{\text{ED}}}

\newcommand{\mue}{\nu_\mu \rightarrow \nu_e}

\newcommand{\mumu}{\nu_\mu \rightarrow \nu_\mu}

\newcommand{\mutau}{\nu_\mu \rightarrow \nu_\tau}
\newcommand{\mutaubar}{\bar{\nu}_\mu \rightarrow \bar{\nu}_\tau}
\newcommand{\pmue}{P(\nu_\mu \rightarrow \nu_e)}
\newcommand{\pmumu}{P(\nu_\mu \rightarrow \nu_\mu)}
\newcommand{\pmuebar}{P(\bar{\nu}_\mu \rightarrow \bar{\nu}_e)}

\newcommand{\chisq}{\Delta\chi^2}

\hypersetup{
  colorlinks,
  citecolor=red,
  linkcolor=blue,
  urlcolor=blue}

\begin{document}

\begin{titlepage}

\renewcommand{\thefootnote}{\alph{footnote}}

\vspace*{1.cm}
\begin{flushright}

\end{flushright}


\renewcommand{\thefootnote}{\fnsymbol{footnote}}
\setcounter{footnote}{0}

{\begin{center}
{\Large  Probing Large Extra Dimension at DUNE using beam tunes 
\\
}
\end{center}}

\renewcommand{\thefootnote}{\alph{footnote}}

\vspace*{.8cm}
\vspace*{.3cm}
{
\begin{center} 
{\sf               Kim Siyeon$^{1,2}$\,\footnote[1]{\makebox[1.cm]{Email:} siyeon@cau.ac.kr},        
                }
   {\sf    Suhyeon Kim$^{1,2}$\,\footnote[2]{\makebox[1.cm]{Email:} sua8897@gmail.com},
                }                                        
        {\sf                 Mehedi Masud$^{1}$\,\footnote[3]{\makebox[1.cm]{Email:} masud@cau.ac.kr},
    }
        {\sf            Juseong Park$^{1,2}$\,\footnote[4]{\makebox[1.cm]{Email:} juseongpark0921@gmail.com}
        }
 		 
\end{center}
}
\vspace*{0cm}
{\it 
\begin{center}

${}^{1}$\ High Energy Physics Center, Chung-Ang University, Seoul 06974, Korea\\
${}^{2}$\ Department of Physics, Chung-Ang University, Seoul 06974, Korea  
\end{center}
}

\vspace*{1.5cm}
%

{\Large 
\bf
 \begin{center}
  Abstract  
\end{center} 
}

The Deep Underground Neutrino Experiment (DUNE) is a leading experiment in neutrino physics
which is presently under construction. DUNE aims to measure the yet unknown parameters in the
three flavor oscillation case which includes discovery of leptonic CP violation, determination
of the neutrino mass hierarchy and measuring the octant of $\theta_{23}$. Additionally, the ancillary goals
of DUNE include probing the subdominant effects induced by possible physics beyond the Standard Model (BSM).
One such new physics scenario is the possible presence of Large Extra Dimension (LED) which can naturally give rise to tiny neutrino masses. 
LED impacts neutrino oscillation through two new parameters, - namely the lightest Dirac mass $m_{0}$ and the radius of the extra dimension $R_{\text{ED}}$ ($< 2$ $\mu$m). 
At the DUNE baseline of 1300 km, the probability seems to be modified more at the higher energy ($\gtrsim 4-5$ GeV) in presence of LED. 
In this work, we attempt to constrain the parameter space of $m_{0}$ and $R_{\text{ED}}$ by performing a statistical analysis of neutrino data simulated at DUNE far detector (FD). 
We illustrate how a combination of the standard low energy (LE) neutrino beam and a medium energy (ME) neutrino beam can take advantage of the relatively large impact of LED at higher energy and improve the constraints. 
In the analysis we also show the role of the individual oscillation channels ($\mue, \mumu, \mutau$), as well as the two neutrino mass hierarchies.
\vspace*{.5cm}

\end{titlepage}

\section{Introduction}
\label{sec:intro}
The phenomenon of neutrino oscillation has been firmly established from the analysis of atmospheric~\cite{Super-Kamiokande:1998kpq} and solar~\cite{SNO:2001kpb} neutrino experiments, which provide an irrefutable evidence that neutrinos are massive particles.  
Various other experiments have confirmed neutrino oscillation till date. 
The global analyses of oscillation data~\cite{10.5281/zenodo.4726908, 
deSalas:2020pgw, nufit_globalfit, Esteban:2024eli, Capozzi:2018ubv} are mostly consistent with the standard three-flavor ($\nu_{e}, \nu_{\mu}, \nu_{\tau}$) neutrino scenario, fully described by three mixing angles $\ta, \tb, \tc$; one CP phase ($\da$) and two mass squared differences $\ldm$, and $\sdm$. 
However, there still remains several unanswered questions in the neutrino sector. 
One such crucial issue is how neutrino gains its tiny mass, since in the Standard Model (SM), neutrinos are massless particles. 
In order to explain the small neutrino masses, in the traditional approach, the SM is extended to include heavy right handed (RH) neutrinos to generate the tiny 
masses of the  neutrinos  via {\it{seesaw}} mechanism~\cite{Minkowski:1977sc, 
Yanagida:1979as, Mohapatra:1979ia, Schechter:1980gr}. 

But there is also an alternative mechanism to generate the small neutrino masses in presence of a new physics model known as Large Extra Dimension (LED). 
The model of LED was originally developed to address another fundamental problem, - the huge gap between the electroweak scale ($M_{\text{EW}} \sim 10^{3}$ GeV) and the Planck scale ($M_{pl} \sim G_{N}^{-1/2} \sim 10^{18}$ GeV) (also known as the {\it{hierarchy problem}})~\cite{Arkani-Hamed:1998jmv, Antoniadis:1998ig, Arkani-Hamed:1998sfv}. 
In this model, the familiar 4-dimensional spacetime forms a hypersurface (known as {\it{brane}}) embedded within a higher dimensional ($4+n$ dimension) spacetime, known as 
the {\it{bulk}}, - $n$ being the number of extra spatial dimensions with size $R$. 
The SM particles (quarks, leptons and gauge bosons) reside in the brane, while gravity can 
propagate in the bulk (full $4+n$ dimensions), - making gravity orders of magnitude weaker in strength compared to other fundamental interactions.  
If $R$ is the radius of the extra dimension, then at a distance $r \lesssim R$, the gravitational force falls off as $r^{-(2+n)}$, while at $r > R$, 
it retains its usual $r^{-2}$ falloff. 
In LED, the electroweak scale is the only fundamental mass scale ($M_{f} \sim M_{\text{EW}} \sim \mathcal{O}(\text{1 TeV})$) and it is related to the Planck scale as
$M_{pl}^{2} \simeq M_{f}^{2+n}R^{n}$. 
This gives the size of the extra spatial dimensions 
as $R \sim 2 \times 10^{(30/n)-17}$ cm.  
One extra spatial dimension ($n=1$) gives $R \sim 10^8$ km, which indicates deviations 
of gravitational force at solar-scale distances and is thus easily excluded from astrophysical observations. 
But $n \gtrsim 2$ suggests deviations of gravitational force at $\mathcal{O}(10^{-1})$ mm 
or below. Thus two or more extra spatial dimensions are still allowed and may 
potentially be detected at future experiments. 

LED can explain the smallness of neutrino mass in an elegant way~\cite{Arkani-Hamed:1998wuz, Dienes:1998sb, Dvali:1999cn, Barbieri:2000mg, Nortier:2020lbs}. 
Since the RH 
neutrinos are singlets under SM gauge group, they can propagate into the bulk. 
The left handed  (LH) neutrinos which are $SU(2)$ doublets in SM are restricted to the brane only. 
Thus the Yukawa couplings between the RH neutrinos and the SM LH neutrinos are 
suppressed by a factor of $M_{\text{EW}}^{2}/M_{pl}$ and this in turn gives the smallness 
of neutrino mass. 
The existence of LED can be probed in neutrino experiments through the effects of 
Kaluza-Klein (KK) excitations which describe the RH neutrinos in the bulk. 
The KK modes describing the higher dimensional RH neutrino fields behave like an infinite tower of sterile neutrinos. 
They induce additional frequencies in the oscillation probabilities  
and generate distortion in a subdominant way. 
The usual practice is to consider an asymmetric space where only one out of $n$ extra spatial dimensions is large, such that there are four spatial dimensions in total. 
Two additional parameters appear in the neutrino oscillation phenomenology in presence of LED, - namely the lightest neutrino mass ($m_{0}$) and the radius/size of the extra dimension  ($\rad$).

Several next-generation long-baseline neutrino experiments are in the pipeline in order to explore the unresolved 
issues in neutrino oscillation physics,- such as, the search for leptonic CP violation (CPV), determination of neutrino mass hierarchy and the correct octant of 
the mixing angle $\tc$ \etc. 
These future experiments include, for \eg,  Deep Underground Neutrino Experiment (DUNE)~\cite{Acciarri:2015uup, DUNE:2016rla}, 
Tokai to Hyper-Kamiokande (T2HK)~\cite{Hyper-KamiokandeProto-:2015xww}, Tokai to Hyper-Kamiokande with a second detector in Korea(T2HKK)~\cite{Hyper-Kamiokande:2016srs}, European Spallation Source $\nu$ Super Beam (ESS$\nu$SB)~\cite{ESSnuSB:2013dql} among others. 
Due to the unprecedented precisions these future 
neutrino facilities are expected to achieve, they are also sensitive to 
the effects of new physics such as the presence of LED. 
In the present work we focus on DUNE and explore its capability to probe the relevant LED parameter space of $(m_{0}-\rad)$. 
DUNE is expected to use the standard low energy (LE) tuned flux (having a peak around $2-3$ GeV and falling quickly at energies $E \gtrsim 4$ GeV) with a total runtime of $13$ years distributed equally between the $\nu$ and $\bar{\nu}$ modes ($6.5$ years + $6.5$ years)~\cite{DUNE:2020ypp}. 
Among the additional fluxes that can be used at DUNE, there is a possibility of using a 
medium energy (ME) beam (also known as the $\nu_{\tau}$-optimized beam) which offers substantial statistics even at energies $E \gtrsim$ $4$ GeV (although at the cost of some loss of statistics around $2-3$ GeV)~\cite{dunefluxes, DUNE:2020ypp}. 
It has already been shown in literature that the combinations of different beams can be optimized in order to improve the sensitivity to new physics parameters and other standard unknowns~\cite{Masud:2017bcf, Rout:2020cxi}. 
In this article we seek to exploit a possible optimized combination of the LE and ME beam in order 
to probe the LED parameters more efficiently. 

In literature, there are several works that analyze the LED parameters at DUNE and other LBL experiments~\cite{Machado:2011jt, DiIura:2014csa, Berryman2016, MINOS:2016vvv, Evans:2017brt, Arguelles:2019xgp, DUNE:2020fgq, 
Forero:2022skg, Khan:2022bcl, Roy:2023dyq, Giarnetti:2024mdt}. 
For \eg, the authors of \cite{Berryman2016, Arguelles:2019xgp, DUNE:2020fgq} study LED parameters at DUNE and further \cite{Berryman2016}  
analyzes the capability of DUNE to distinguish between different new physics scenarios such as that of LED and light sterile neutrinos.  
The authors of \cite{Forero:2022skg} have done an analysis of LED with the available data from MINOS/MINOS+~\cite{MINOS:2017cae} and Daya Bay~\cite{DayaBay:2012fng} experiments and have also projected  
the future constraints from direct neutrino mass measurement experiments like KATRIN~\cite{KATRIN:2001ttj}. 
The author of \cite{Roy:2023dyq} has probed the LED parameters using both the charged-current and neutral current interactions at DUNE and also compared the results by estimating sensitivities from other LBL experiments such as T2HK and ESS$\nu$SB. 
Most recently, the authors of \cite{Giarnetti:2024mdt} have analyzed the sensitivities to LED parameters at DUNE when using a high-energy tuned beam.
There are three distinct aspects in which the present work goes beyond the other existing works on LED at LBL experiments.
\begin{itemize}
\item In the standard analyses of LED at DUNE, the beam used is generally the standard low energy (LE)-tuned beam. 
In the present work we exploit the higher energy effects of LED more efficiently and use the medium energy (ME) tuned beam ($\nu_{\tau}$-optimised beam) in conjunction with the LE beam in order to improve the constraints on LED parameter space. 
We would like to emphasize that our study estimates an optimized combination of runtimes shared between the LE and ME beam so that we utilise the statistics from both the low-energy and higher-energy bins~\footnote{This is in contrast to the study conducted in \cite{Giarnetti:2024mdt} where only one high energy beam was used to constrain LED.}.

\item We include the contributions from $\mutau$ channel (in addition to $\mue$ and $\mumu$) in our analyses and also discuss 
the individual roles of all three channels in constraining LED. 

\item We use the most recent GLoBES configurations by the DUNE collaboration~\cite{DUNE:2021cuw}. 
This has a runtime of 13 years with 624 kt.MW.yr.\ exposure.
\end{itemize}

The article is organized as follows. 
We start with a brief overview of the relevant theoretical basics of LED in Sec.\ \ref{sec:theory} in the context of neutrino oscillation,
 and then follow it up with a probability level discussion 
in Sec.\ \ref{sec:prob}. 
We then discuss about the different beam tunes 
and the corresponding event spectra in Sec.\ \ref{sec:event}. 
This is followed by an outline of the statistical procedure adopted in Sec.\ \ref{sec:chisq}, followed by our main sensitivity results in Sec.\ \ref{sec:results}. 
Finally we conclude in Sec.\ \ref{sec:conclusion}.

\section{Theoretical basics of LED}
\label{sec:theory}
In presence of LED, the fields carrying charge under the SM gauge symmetries are confined to the familiar 4d spacetime (brane) and the fields that are singlets under SM gauge symmetries can propagate into the (4+N) dimensional spacetime (bulk). 
Following the usual treatment of LED models in neutrino oscillation phenomenology~\cite{Davoudiasl2002, Esmaili:2014esa, Machado:2011jt, Machado:2011kt, Basto-Gonzalez:2012nel, Girardi:2014gna, Rodejohann:2014eka, Berryman2016, Carena:2017qhd, Stenico:2018jpl, Basto-Gonzalez:2021aus, Forero:2022skg, Roy:2023dyq} we consider 
effectively a (4+1) dimensional spacetime, such that one extra dimension is compactified 
on a circle with radius $\rad$, which is much larger than the other extra dimensions. 
We consider three right handed neutrino fields $\nu_{\alpha R}$ (lying in the 5d spacetime) associated with the three 
left-handed neutrino fields $\nu_{\alpha L} (\alpha = e,\mu,\tau)$ residing in the 4d spacetime. 
The $\nu_{\alpha R}$ can be expressed as infinite number of Kaluza-Klein modes after imposing the periodic 
boundary conditions due to the compactification of the fifth dimension. 
The relevant mass term of the lagrangian after electroweak symmetry breaking becomes~\cite{Machado:2011jt},
\begin{equation}
\mathcal{L}_{\text{mass}} = 
\sum_{\alpha, \beta}m_{\alpha\beta}^{D}\bigg[
\bar{\nu}_{\alpha L}^{(0)}\nu_{\beta R}^{(0)}
+ \sqrt{2}\sum_{n=1}^{\infty}\bar{\nu}_{\alpha L}^{(0)}\nu_{\beta R}^{(n)})
\bigg]
+ \sum_{\alpha}\sum_{n=1}^{\infty}
\frac{n}{\rad}\bar{\nu}_{\alpha L}^{(n)}\nu_{\alpha R}^{(n)} + h.c.,
\label{eq:lagrangian}
\end{equation}
where $\alpha, \beta = e, \mu,\tau$; $m_{\alpha\beta}^{D}$ is the Dirac mass matrix. 
$\nu_{\alpha R}^{(0)}, \nu_{\alpha R}^{(n)}, \nu_{\alpha L}^{(n)}$ are the linear 
combinations of the bulk fermion fields that couple to the SM neutrino fields $\nu_{\alpha L}^{(0)}$, and the index $n$ denotes the KK mode. 
After diagonalizing the Dirac mass matrix $m^{D}$ using U and R such that  $U^{\dagger}m^{D}R = \text{diag}(m_{1}^{D}, m_{2}^{D}, m_{3}^{D})$, the 
mass term becomes,
\begin{equation}
\mathcal{L}_{\text{mass}} = \sum_{i=1}^{3}\bar{\nu}_{i L}^{\prime}M_{i}\nu_{i R}^{\prime} + h.c.,
\label{eq:lagrangian1}
\end{equation}
where $M_{i}$ is an infinite dimensional matrix:
\begin{equation}
M_{i} = \frac{1}{\rad}\begin{pmatrix}
m_{i}^{D}\rad & 0 & 0 &...& 0\\
\sqrt{2}m_{i}^{D}\rad & 1 & 0 & ... & 0\\
\sqrt{2}m_{i}^{D}\rad & 0 & 2 & ...& 0\\
.. & .. & .. & ..& ..\\
.. & .. & .. & ..& ..\\
\sqrt{2}m_{i}^{D}\rad & 0 & 0 & ...& n
\end{pmatrix}.
\label{eq:M}
\end{equation}
The new basis in Eq.\ \ref{eq:lagrangian1} are given by,
\begin{align}
\nu_{\alpha L}^{(0)} = \sum_{i}U_{\alpha i}\nu_{i L}^{\prime(0)}, 
\text{\qquad}
\nu_{\alpha R}^{(0)} = \sum_{i}R_{\alpha i}\nu_{i R}^{\prime(0)} \nonumber \\
\nu_{\alpha L,R}^{(n)} = \sum_{i}R_{\alpha i}\nu_{i L, R}^{\prime(n)} 
\text{\qquad with $n>0$}.
\label{eq:basis}
\end{align}
The true mass is obtained after diagonalisation of the infinite-dimensional 
matrix $M_{i}$ and the subsequent mixing of the three active neutrinos is given by,
\begin{equation}
\nu_{\alpha L} = \sum_{i=1}^{3}U_{\alpha i}\sum_{n=0}^{\infty}V_{i}^{(n)}\nu_{i L}^{\prime (n)},
\label{eq:mixing}
\end{equation}
where U is the standard $3 \times 3$ unitary leptonic mixing matrix. 
$\nu_{i L}^{\prime (n)}$ is a neutrino field with mass $m_{i}^{n} = \lambda_{i}^{n}/\rad$ where 
$\lambda_{i}^{n}$ are the solutions of the eigenvalue equation,
\begin{equation}
\lambda_{i}^{n} - \pi(m_{i}^{D}\rad)^{2}\cot(\pi\lambda_{i}^{n}) = 0.
\label{eq:evalue}
\end{equation}
$m_{i}^{D} (i=1,2,3)$ are the three eigenvalues of the Dirac neutrino mass matrix, which are 
strongly suppressed by the LED volume factor equal to $(M_{f}/M_{pl})$~\cite{Dvali:1999cn}. 
The eigenvalue equation has infinite number of solutions for $\lambda_{i}^{n}$ in the 
interval $[n,n+1/2]$ (for $n=0,1,2,...\infty$). This implies,
\begin{equation}
\frac{n}{\rad} < m_{i}^{n} < \frac{n+1/2}{\rad} \text{\qquad for $n=0,1,2,..\infty$}.
\label{eq:mi_limit}
\end{equation}
The elements of the mixing matrix $V$ in Eq.\ \ref{eq:mixing} are given by~\cite{Dienes:1998sb, Dvali:1999cn, Mohapatra:2000wn},
\begin{equation}
(V_{i}^{n})^{2} = \frac{2}{1+\pi^{2}(m_{i}^{D}\rad)^{2} + (m_{i}^{n}/m_{i}^{D})^{2}}.
\label{eq:V}
\end{equation}
Assuming LED to be a perturbative effect ($\xi_{i} = m_{i}^{D}\rad << 1$) on top of the standard three neutrino oscillation, 
it is possible to solve 
the eigenvalue equation analytically to obtain the following approximate expressions~\cite{Davoudiasl2002}.
\begin{align}
&m_{i}^{0} \simeq m_{i}^{D}\bigg[1-\frac{\pi^{2}}{6}\xi_{i}^{2} + ... \bigg], 
\text{\qquad}
V_{i}^{0} = 1-\frac{\pi^{2}}{6}\xi_{i}^{2} + ..., \nonumber \\
&m_{i}^{n} \simeq \frac{n}{\rad}\bigg[1 + \frac{\xi_{i}^{2}}{n^{2}} + ... \bigg], 
\text{\qquad}
V_{i}^{n} = \sqrt{2}\frac{\xi_{i}}{n}\bigg[1 - \frac{3}{2}\frac{\xi_{i}^{2}}{n^{2}} + ... \bigg] 
\text{\qquad for $n>0$}.
\label{eq:perturbation}
\end{align}
The neutrino oscillation probability in presence of LED from a flavour $\alpha$ to a flavour $\beta$ is given by,
\begin{equation}
P_{\nu_{\alpha} \to \nu_{\beta}}^{\text{LED}} = \bigg|
\sum_{j=1}^{3}\sum_{n=0}^{\infty}U_{\alpha j}^{*}U_{\beta j}(V_{j}^{n})^{2}
\text{exp}\bigg(-i\frac{(m_{j}^{n})^{2}L}{2E} \bigg)
\bigg|^{2},
\label{eq:prob}
\end{equation}
where $E$ is the neutrino energy and $L$ is the baseline length.
Expanding Eq.\ \ref{eq:prob}, it can be seen that $P_{\nu_{\alpha} \to \nu_{\beta}}^{\text{LED}}$ contains the interference phases,
\begin{equation} 
\phi_{jk}^{mn} = \bigg[(m_{j}^{m})^{2} - (m_{k}^{n})^{2}\bigg]\frac{L}{2E} 
= \Delta (m_{jk}^{mn})^{2}\frac{L}{2E}
\text{\quad $(j,k=1,2,3$; $m,n=0,1,2,..\infty)$}.
\label{eq:phi}
\end{equation}
Further, each of these mass-induced interference terms are also proportional to the LED mixing term $(V_{j}^{m})^{2}(V_{k}^{n})^{2}$ (in addition to the mixing induced by the standard PMNS matrix $U$). 
The phase differences can be seen as originating from the following three types of interferences,
\begin{itemize}
\item \textit{Interferences among standard neutrinos (\ie, $0$-mode KK neutrinos)}:
In this case the corresponding phase and the LED mixing terms in leading orders are (using Eqs.\ \ref{eq:perturbation}),
\begin{align}
 &\phi_{jk}^{00} = \Delta (m_{jk}^{00})^{2}\frac{L}{2E}, \nonumber \\ 
 &(V_{j}^{0})^{2}(V_{k}^{0})^{2} \simeq 1-\frac{\pi^{2}}{3}(\xi_{j}^{2}+\xi_{k}^{2})
 \text{\quad $(j,k=1,2,3)$}.
 \label{eq:phi_00}
 \end{align}
 The expressions in Eq.\ \ref{eq:phi_00} contribute to two kinds of terms in the probability expression of Eq.\ \ref{eq:prob}. 
 The terms proportional to $exp(-i\phi_{jk}^{00})$ can be recognized roughly as the contribution from the standard oscillation probability in the three neutrino case. 
 On the other hand, the terms proportional to $-\frac{\pi^{2}}{3}(\xi_{j}^{2}+\xi_{k}^{2})exp(-i\phi_{jk}^{00})$ give the 
 correction to the probability due to the presence of LED. 
 Note that this correction has a negative sign implying that the presence of LED tends to decrease the magnitude of oscillation probability. 
 \item \textit{Interferences between the standard and $(n > 0)$ KK mode neutrinos}:
 In this case, the corresponding phase and the mixing in leading orders can be derived as,
 \begin{align}
& \phi_{jk}^{0n} = \Delta (m_{jk}^{0n})^{2}\frac{L}{2E} \simeq -\frac{n^{2}}{\rad^{2}}\frac{L}{2E}, \nonumber \\
&(V_{j}^{0})^{2}(V_{k}^{n})^{2} \simeq \frac{2}{n^{2}}\xi_{k}^{2} 
 \text{\quad $(j,k=1,2,3; n > 0)$}.
 \label{eq:phi_0n}
 \end{align}
 For a typical value of $\rad = 0.5$ $\mu$m $\simeq 2.5 \text{ eV}^{-1}$, the mass-squared difference
 corresponding to the lowest mode ($n=1$) KK neutrino is 
 $\Delta (m_{jk}^{01})^{2} \simeq -0.16 \text{ eV}^{2}$, such that $|\Delta (m_{jk}^{01})^{2}|$ is orders of magnitude larger than the standard mass squared differences $|\ldm|$ ($\sim 10^{-3} \text{ eV}^{2}$) and $\sdm$ ($\sim 10^{-5} \text{ eV}^{2}$). 
 The mass-squared differences for higher KK modes are even higher in magnitude ($\propto n^{2}$).
 Thus for the typical values of $L \sim \mathcal{O}(10^{3})$ km and $E \lesssim \mathcal{O}(10)$ GeV, 
 the interference terms with phases $\phi_{jk}^{0n}$ ($n > 0$) get largely averaged out.  
 Further, Eq.\ \ref{eq:phi_0n} shows that the corresponding interference terms are suppressed by the small perturbative term proportional to $\xi_{k}^{2}$ ($k=1,2,3$) as well, and this suppression gets even stronger for higher KK modes ($\propto n^{-2}$). 
 \item  \textit{Interferences among the $(n > 0)$ KK mode neutrinos}:
 The relevant phase and the mixing factors can be shown as (in leading orders),
 \begin{align}
 &\phi_{jk}^{mn} \simeq \frac{1}{\rad^{2}}\bigg[
 m^{2}-n^{2} + 2(\xi_{j}^{2}-\xi_{k}^{2})
 \bigg] \frac{L}{2E}, \nonumber \\
 &(V_{j}^{m})^{2}(V_{k}^{n})^{2} \simeq \frac{4}{m^{2}n^{2}}\xi_{j}^{2}\xi_{k}^{2} 
 \text{\quad $(j,k=1,2,3; m,n > 0)$}.
 \label{eq:phi_mn}
  \end{align}
 The phases in Eqs.\ \ref{eq:phi_mn} are as large as $\phi_{jk}^{0n}$ (in Eqs.\ \ref{eq:phi_0n}) in magnitudes and the interference terms
 are increasingly suppressed by factors $\sim \mathcal{O}(\xi^{4})$. Hence these interference terms provide very little contribution 
 to the probability in Eq.\ \ref{eq:prob}.
 Both Eqs.\ \ref{eq:phi_0n} and \ref{eq:phi_mn} justify our choice of considering only upto $n=2$ KK neutrinos in the analysis.
\end{itemize}

In presence of matter, the oscillation probability gets modified after taking into account the 
charged and neutral current interactions experienced by neutrino. 
In matter the probability is determined by the following evolution equation~\cite{Berryman2016}.
\begin{equation}
i\frac{d}{dt}\nu_{jL}^{\prime} = \bigg[
\frac{1}{2E}M_{j}^{\dagger}M_{j}\nu_{jL}^{\prime} 
+ 
\sum_{k=1}^{3}\begin{pmatrix}
Y_{jk} & 0_{1 \times n} \\
0_{n \times 1} & 0_{n \times n}
\end{pmatrix}
\nu_{jL}^{\prime}
\bigg]_{n \to \infty} 
\text{ with }
Y_{jk} = \sum_{\alpha=e,\mu,\tau}U_{\alpha j}^{*}U_{\alpha k}
\big(
\delta_{\alpha e}V_{\text{CC}} + V_{\text{NC}}
\big).
\label{eq:prob_matter}
\end{equation}
The basis are given by Eqs.\ \ref{eq:basis}. 
The charged and neutral current potentials are given by $V_{\text{CC}} = \sqrt{2}G_{F}n_{e}$ 
and $V_{\text{NC}} = -(1/\sqrt{2})G_{F}n_{n}$, where $n_{e} (n_{n})$ are the electron (nucleon) 
number density along the path of neutrino propagation. 

We implement the physics of LED and the oscillation probability in the General Long Baseline Experiment Simulator (GLoBES)~\cite{Huber:2004ka,Huber:2007ji} 
(see Appendix for details) and proceed in the following 
way. 
Since the impact of LED is perturbative in nature, we require the zero-th KK mode to correspond 
to the standard mass-squared differences~\cite{Basto-Gonzalez:2021aus}:
\begin{equation}
\Delta m^{2}_{kj} = (m_{k}^{0})^{2} - (m_{j}^{0})^{2}
\label{eq:kk_0}
\end{equation}
The experimentally measured two standard mass-squared differences $\sdm$ and $\ldm$ 
help us (using Eqs.\  \ref{eq:evalue} and \ref{eq:kk_0}) to fix two of the independent 
Dirac masses $m_{i}^{D}$. 
Choosing the lightest Dirac mass $m_{0}$ to be a free parameter of the LED model, all 
three Dirac masses $m_{i}^{D}$ can then be determined. 
Clearly for the NH scenario, $m_{0} = m_{1}^{D}$ and for IH $m_{0} = m_{3}^{D}$. 
Once all the three Dirac masses are estimated, we use Eq.\ \ref{eq:evalue} to obtain the  
 neutrino masses $m_{i}^{n}$ for higher KK modes ($n > 0$). 
 We then use Eq.\ \ref{eq:prob} to determine the oscillation probability with $m_{0}$ and $\rad$ 
 as the additional model parameters in addition to the six standard oscillation parameters. 
 We have considered upto two KK modes (\ie, upto $n = 2$) and have checked that the changes in oscillation probability is  negligible for higher KK modes.  
 
It needs to be mentioned that there exists a phenomenological constraint on $\rad$~\cite{Forero:2022skg}. 
From Eq.\ \ref{eq:mi_limit}, for the zero-mode mass $\rad m_{k} < 1/2$ ($k=1,2,3$). 
But since,
\begin{equation}
 m_{k}^{2} = m_{j}^{2} + \Delta m_{kj}^{2} \geq \Delta m_{kj}^{2}, 
 \end{equation}
we can write,
\begin{equation}
\rad \leq \frac{1}{2\sqrt{\Delta m_{kj}^{2}}} \text{\qquad $k,j = 1,2,3$ and $k \neq j$}
\end{equation}
The largest such mass-squared difference is the atmospheric mass-squared difference and 
is around $2.5 \times 10^{-3} \text{ eV}^{2}$ (which is $\Delta m^{2}_{31}$ for NH, and $
\Delta m^{2}_{23}$ for IH). 
Thus the physically allowed values of $\rad$ should satisfy,
\begin{equation}
\rad \leq \frac{1}{2\sqrt{2.5 \times 10^{-3}}} \text{ eV}^{-1} \lesssim 2 \text{ $\mu$m}.
\label{eq:R_pheno_bound}
\end{equation}
\section{Probabilities in presence of LED}
\label{sec:prob}
 \begin{table}[h]
\centering
\begin{tabular}{| c | c | c | c |}
\hline
&&&\\
Parameter & Best-fit-value & 3$\sigma$ interval & $1\sigma$ uncertainty  \\
&&&\\
\hline
&&&\\
$\theta_{12}$ [Deg.]             & 34.3                    &  31.4 - 37.4   &  2.9\% \\
$\theta_{13}$ (NH) [Deg.]    & 8.53  &  8.13  -  8.92   &  1.5\% \\
$\theta_{13}$ (IH) [Deg.]    & 8.58  &  8.17  -  8.96   &  1.5\% \\
$\theta_{23}$ (NH) [Deg.]        & 49.3      &  41.2  - 51.4    &  3.5\% \\
$\theta_{23}$ (IH) [Deg.]        & 49.5                     &  41.1  - 51.2    &  3.5\% \\
$\sdm$ [$\text{eV}^2$]  & $7.5 \times 10^{-5}$  &  [6.94 - 8.14]$\times 10^{-5}$  &  2.7\% \\
$\ldm$ (NH) [$\text{eV}^2$] & $2.55 \times 10^{-3}$   &  [2.47 - 2.63] $\times 10^{-3}$ &  1.2\% \\
$\ldm$ (IH) [$\text{eV}^2$] & $-2.45 \times 10^{-3}$  & -[2.37 - 2.53]$\times 10^{-3}$  &  1.2\% \\
$\delta_{13}$ (NH) [Rad.]   & $-0.92\pi$   & $[-\pi, -0.01\pi]  \cup [0.71\pi, \pi]$ &  $-$ \\
$\delta_{13}$ (IH) [Rad.]   & $-0.42\pi$   & $[-0.89\pi, -0.04\pi]$   & $-$  \\\
&&&\\
\hline
\end{tabular}
\caption{\footnotesize{\label{tab:parameters}
 The values of the standard oscillation parameters and their uncertainties used in our study. 
 The values were taken from the global fit analysis in \cite{10.5281/zenodo.4726908, deSalas:2020pgw}. 
If the $3\sigma$ upper and lower limit of a parameter is $x_{u}$ and $x_{l}$ respectively, the $1\sigma$  uncertainty is $(x_{u}-x_{l})/3(x_{u}+x_{l})\%$~\cite{DUNE:2020ypp}. 
}}
\end{table}
\begin{figure}[h]
    \centering
        \includegraphics[scale=0.6]{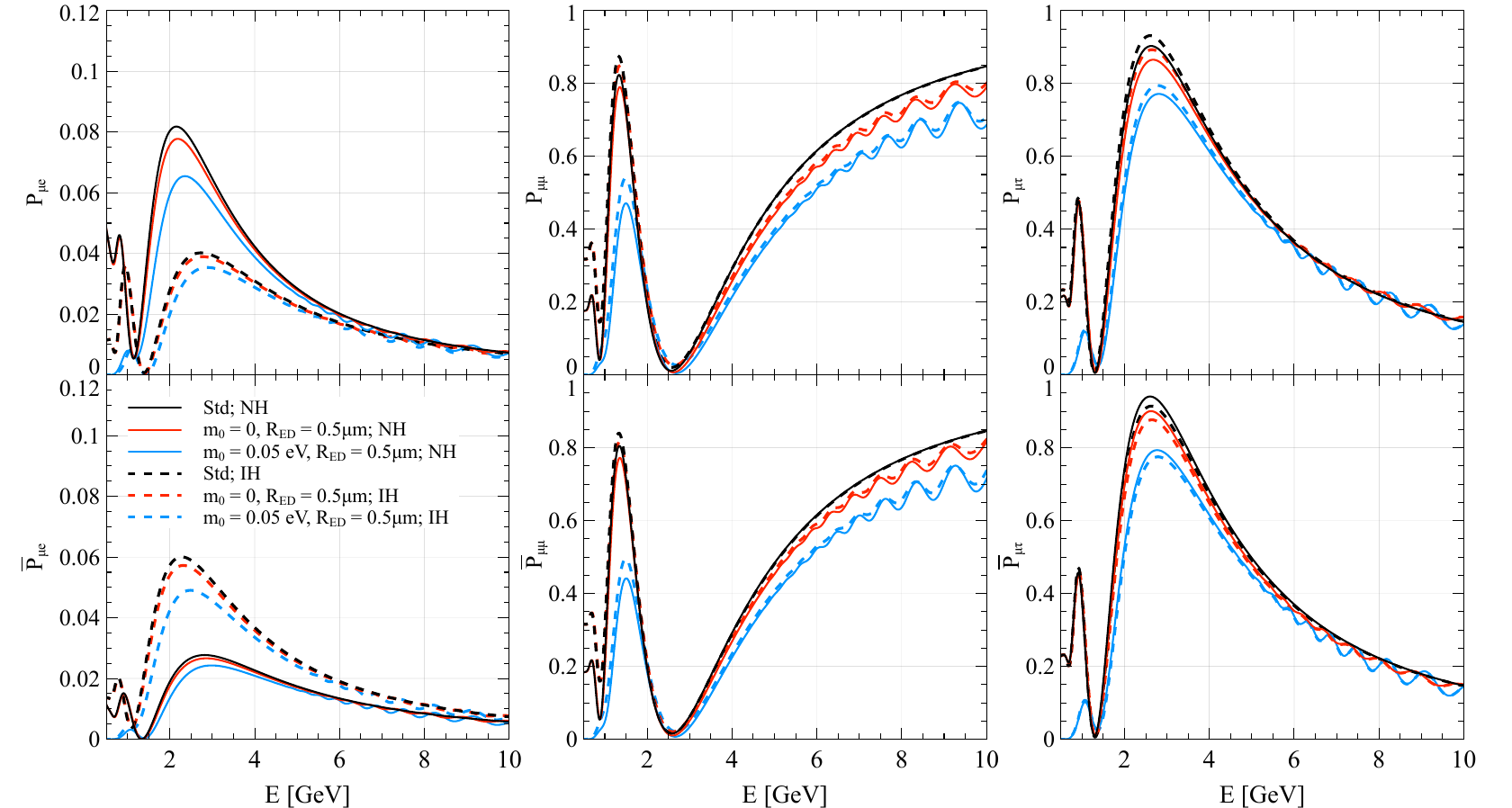}
    \caption{
        \footnotesize{
        This shows the probabilities as functions of energy ($E$) for both standard (std.) scenario and in presence of LED at the DUNE FD baseline of 1300 km. 
        The three columns show the cases of the three channels $\mue, \mumu$ 
        and $\mutau$, while the top (bottom) row depict the case of neutrinos (antineutrinos). 
        The black curves correspond to the std. case while the red and blue curves illustrate 
        the presence of LED (with two different values for the LED parameter $m_{0}$ as shown in the figure legend). The LED compactification radius $\rad$ is chosen as 0.5 $\mu$m for the LED case. The values of all standard oscillation parameters were chosen from Table \ref{tab:parameters}.   The solid (dashed) curves correspond to NH (IH).   
   } }
    \label{fig:prob}
\end{figure}
In Fig.\ \ref{fig:prob} we plot $P(\accentset{(-)}{\nu}_{\mu} \to \accentset{(-)}{\nu}_{e})$, 
$P(\accentset{(-)}{\nu}_{\mu} \to \accentset{(-)}{\nu}_{\mu})$ and $P(\accentset{(-)}{\nu}_{\mu} \to \accentset{(-)}{\nu}_{\tau})$ as a function of $E$ for both the standard (std) and LED case 
at the DUNE baseline of 1300 km. 
The black curves correspond to the standard case. 
The red (blue) curves show the LED case 
with the LED parameter as $m_{0} = 0$ ($m_{0} = 0.05 \text{ eV}$). 
The compactification radius $\rad$ is 0.5 $\mu$m for all the LED cases. 
The solid (dahsed) curves in a panel depict the case of NH (IH), and the top (bottom) row 
shows the case of neutrino (antineutrino). 
We note that the presence of LED has a two-fold effect:
Firstly they introduce new frequencies (driven by the interferences 
between the standard \ie, zero mode masses and \ie, $n > 0$ KK mode masses) 
leading to small wiggles in the probability~\footnote{Presence of the $n$-th KK mode introduces mass-squared difference $\sim \mathcal{O}(n^2 \times 0.16) \text{ eV}^{2}$ (see Eq.\ \ref{eq:phi_0n}), - which 
introduces rapid small oscillations on top of the standard oscillations in the probability spectra. 
But such fast oscillations are not 
observable due to the finite energy resolution of the detectors. 
So, in practice it is more realistic to show 
the somewhat smoothened oscillations after applying the low pass filter in GLoBES with a filter value equal to 125 MeV, - same 
as the bin widths we will use later for the event spectra and subsequent statistical analysis. 
But even after 
such a smoothening, the secondary oscillations due to the KK modes show up as small wiggles 
which can be observed in principle.}. 
Secondly, LED decreases the magnitude of the probability (since the SM neutrinos can oscillate 
into the KK modes). This reduction in magnitude is more for higher values of $m_{0}$.
Both of these effects are more prominent at higher energies ($E \gtrsim 6$ GeV).

The overall reduction of probability is mainly due to the correction introduced by terms proportional to
$(V_{j}^{0})^{2}(V_{k}^{0})^{2}$ (see Eq.\ \ref{eq:phi_00}). 
The phase differences $\phi_{jk}^{0n}$ (with $(j,k = 1,2,3); (n > 0)$) introduced by the $n>0$ KK modes (see Eq.\ \ref{eq:phi_0n} and the 
relevant discussions) are very large compared to the ones induced by the standard ($n=0$ KK mode) 
mass-squared differences.
Thus the interference terms involving the phase differences $\phi_{jk}^{0n} = \Delta (m_{jk}^{0n})^{2}L/2E$ get averaged out. 
At low $E$, clearly such terms have high frequencies leading to a complete averaging out. 
But the frequencies decrease slightly at higher energies which leads to partial non-averaging of the lowest KK mode. Thus the LED effect becomes more apparent at higher energies giving rise to small oscillations/wiggles in the probability spectra. 
Detailed analytical explanations of such reduction in probability in presence of LED have also been 
discussed previously, for \eg, in \cite{Dvali:1999cn}.

The impacts of LED are clearly more prominent for the $\mumu$ channel (second column).  
The results in this channel neither depend upon $\numu \to \numu$ or $\numubar \to \numubar$ channel nor on the neutrino mass hierarchy.
On the other hand, For $\pmue$ ($\pmuebar$), when going to from IH from NH, the values of probabilities for both std.\ and LED case are reduced (increased) considerably by roughly a factor $2-3$. 
Additionally we also see a slight increase (decrease) of probability in the $\mutau$ ($\mutaubar$) channel and when changing mass hierarchy from NH to IH.  


\section{Beam tunes and event spectra}
\label{sec:event}
\begin{figure}[htb]
    \centering
    \includegraphics[scale=0.9]{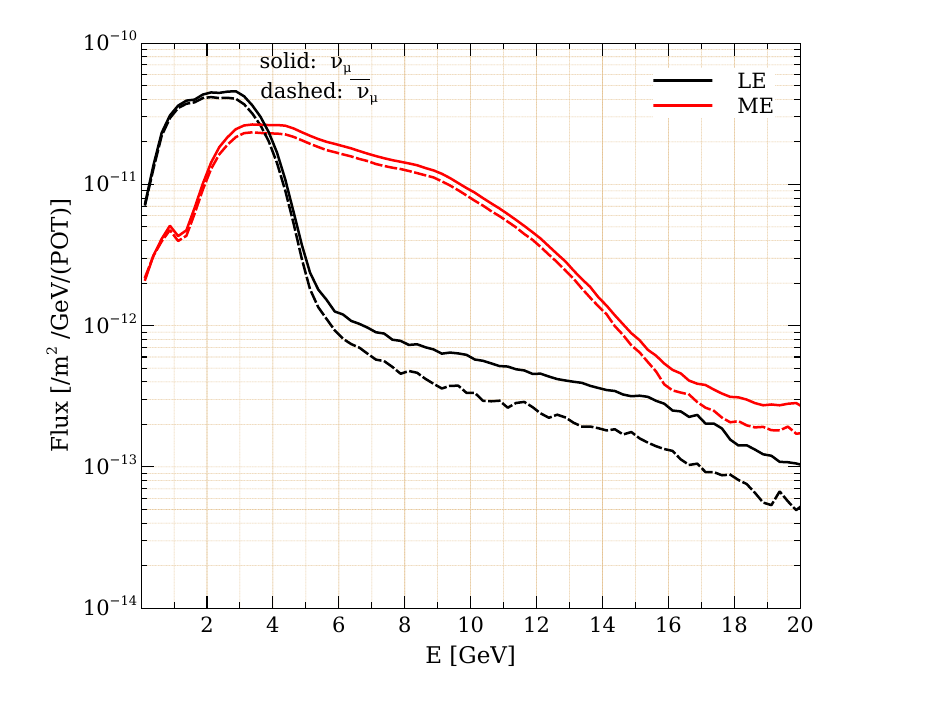}
    \caption{\footnotesize{This shows the standard low energy (LE) muon neutrino beam (black) and the ($\nutau$-optimized) medium energy-tuned muon neutrino beam (red) at DUNE~\cite{DUNE:2020ypp}. 
The solid (dashed) curves correspond to the beam in $\nu$ ($\bar{\nu}$) mode.}}
    \label{fig:flux}
\end{figure}
Our goal is to probe the difference between the standard and LED scenario in order to explore 
the LED parameter space.  
For generating the events in the two cases,  we carry out simulations using GLoBES.  
The most recent configuration files from the Technical Design Report (TDR) of DUNE~\cite{DUNE:2021cuw} have been used in our simulations. 
DUNE consists of an on-axis $40$ kiloton (kt) liquid argon FD housed at the Homestake Mine in South Dakota with a baseline of $L = 1300$ km. 
A near detector (ND)  with target mass $0.067$ kt will be installed at a  baseline $0.570$ km at the Fermi National Accelerator Laboratory (FNAL), in Batavia, Illinois. 
 We use the following broad-band beam tunes : 
\begin{itemize}
\item The standard low energy (LE) beam tune used in DUNE TDR~\cite{DUNE:2020ypp}.
\item The medium energy (ME) beam tune optimized for $\nu_{\tau}$ appearance~\cite{DUNE:2020ypp, dunefluxes}.
\end{itemize}
Both the beams are produced by a $120$ GeV proton beam impinging on a graphite target (delivering $1.1 \times 10^{21}$ protons on target per year) and are obtained from G4LBNF, a GEANT4 based simulation~\cite{Agostinelli:2002hh,Allison:2006ve} of the long baseline neutrino facility (LBNF) beamline~\cite{DUNE:2020ypp}. 
The hadrons produced in the graphite target are then focussed using three magnetic horns operated with 
$300$ kA current and are allowed to decay in a helium-filled decay pipe of length 194 m to produce the LE flux. 
The higher energy tuned ME flux is simulated by 
replacing the three magnetic horns with two NuMI-like parabolic horns with the second horn starting $17.5$ m downstream from the start of the first horn. 
For both the fluxes, the focusing horns can be operated in forward and reverse current configurations to produce beams in $\nu$ 
and $\bar{\nu}$ modes respectively. 
 The two beam tunes used in our study are shown in Fig.\ \ref{fig:flux}. 
 The LE beam (black curve) peaks relatively sharply around $\sim 2-3$ GeV and falls rapidly beyond roughly 4 GeV. 
The ME flux on the other hand peaks around $\sim 3-5$ GeV (with a \textit{broad} peak) and falls much slowly thereby retaining 
significantly higher flux of neutrinos (antineutrinos) compared to the LE beam at $E \gtrsim 4$ GeV. 
\begin{figure}[htb]
    \centering
    \includegraphics[scale=0.57]{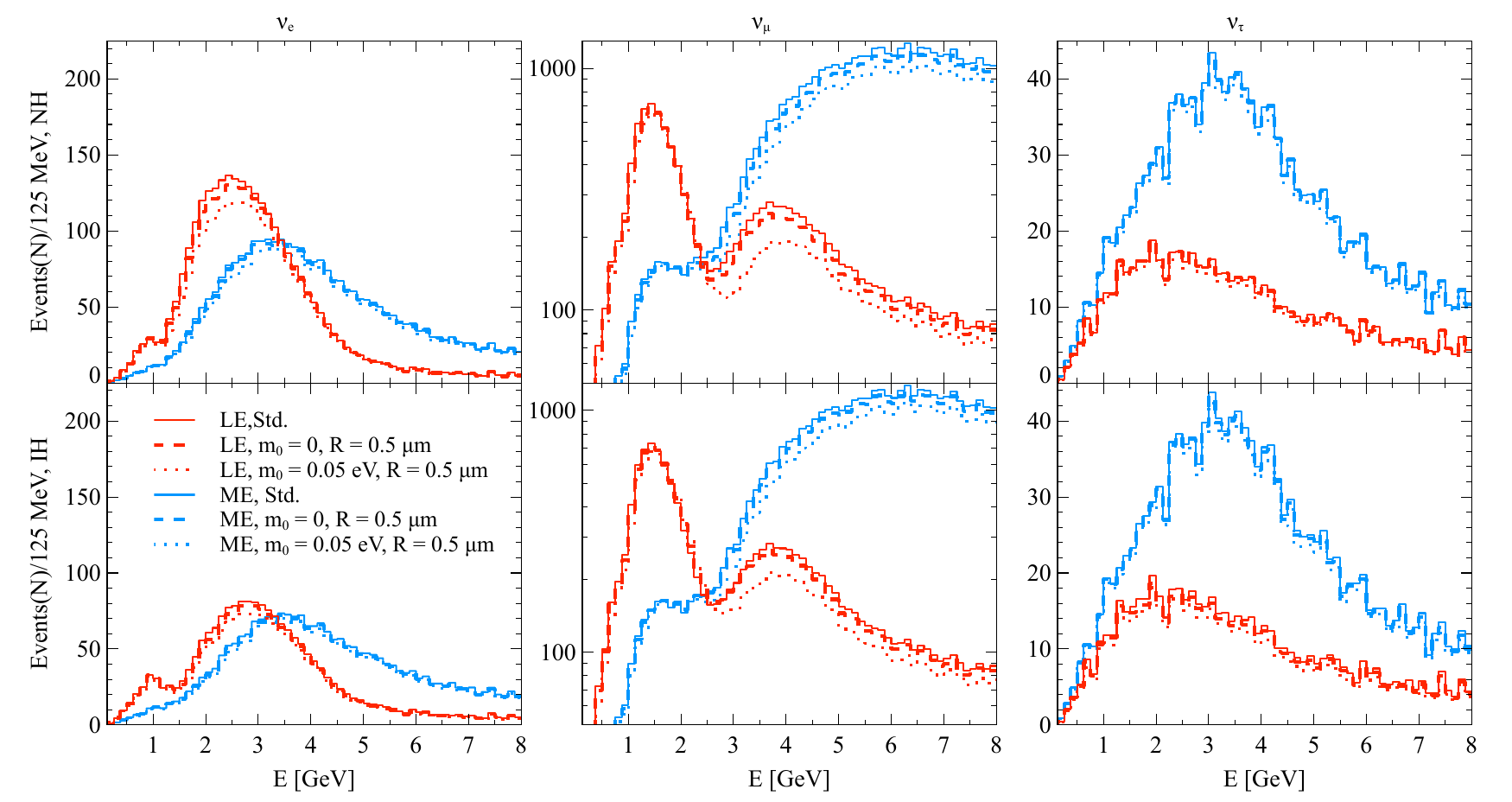}
    \caption{\footnotesize{This shows the event spectra generated at DUNE FD, using either the standard LE beam (red) or the ME beam (blue) with a total of 13 yrs. of runtime (6.5 yrs. each in $\nu$ and $\bar{\nu}$ mode).
    The top (bottom) row corresponds to NH (IH). 
    The solid curves correspond to the standard three neutrino scenario, while the thick dashed and the thin dotted curves correspond to the LED case with different sets of the choice of parameters $(m_{0},\rad)$ as shown in the legend.    
    The three columns correspond to the $\nue$-like, $\numu$-like and $\nutau$-like events respectively. Note that the vertical axes of the second column ($\numu$ events) are in 
    logarithmic scale for ease of comparison between the events generated by LE and ME beams. }}
    \label{fig:event}
\end{figure}
In Fig.\ \ref{fig:event}, we show the event spectra for both std. case and in presence of LED at DUNE FD generated using either the LE beam (red) or ME beam (blue) with a total runtime of 13 years (6.5 yrs.\ each in $\nu$ and $\bar{\nu}$ mode) 
corresponding to an exposure of 624 kt.MW.yrs. 
As expected from Fig.\ \ref{fig:prob} and the relevant discussions in Sec.\ \ref{sec:prob}, 
the events show a decrease in presence of LED.  
and the degree of this decrease is more
for $\numu$-like events. 
Though the event spectra simulated using the LE beam (red) is higher at lower energies, 
the spectra using the ME beam (blue) takes over at higher energies ($E \gtrsim 3.5$ GeV 
for $\nue$ spectra, and $E \gtrsim 2.5$ GeV for $\numu$ spectra). 
This observation is especially significant for the $\numu$ events, - increasing the number of events by more than an order of magnitude when using the ME beam. 
 For \eg, the number of $\numu$ events (both for the std.\ case and in presence of LED) with LE beam lie
  approximately within 
 $40-80$  around 8 GeV. 
 But with the ME beam, these numbers can shoot up to around $900-1000$ at 8 GeV. 
 Since the value of $\pmumu$ is high (close to 1) at higher energies (see Fig.\ \ref{fig:prob}), 
 the use of the ME beam (which is expected to offer more statistics at high energies, - see Fig.\ \ref{fig:flux}) is thus particularly advantageous for the $\mumu$ channel. 
 Such a huge statistics at energies beyond 4 GeV is thus expected to offer more sensitivities 
 in probing the effects of LED. 
For the $\mue$ channel, the number of events $N$ with the LE beam at $E \gtrsim 4$ GeV is very small (for \eg, $N \simeq 5-10$ around 8 GeV for NH: left-top panel of Fig.\ \ref{fig:event}) owing to the sharply falling LE flux. 
But ME beam offers high statistics at higher energy, - thus increasing the number of events to 
around 20-60 in the range 4-8 GeV for NH. 
Since the ME beam is already optimized for $\mutau$ appearance events, the number of 
$\nutau$ events is significantly higher compared to the case when LE beam is used. 
For \eg, in the NH case (top-right panel of Fig.\ \ref{fig:event}) the number of $\nutau$ events at the peak of the spectrum when using LE is around 16 ($E \simeq 2$ GeV). 
But using ME beam, the number of $\nutau$ events at the peak is around 40 ($E \simeq 3$ GeV). 

\begin{figure}[htb]
    \centering
    \includegraphics[scale=0.7]{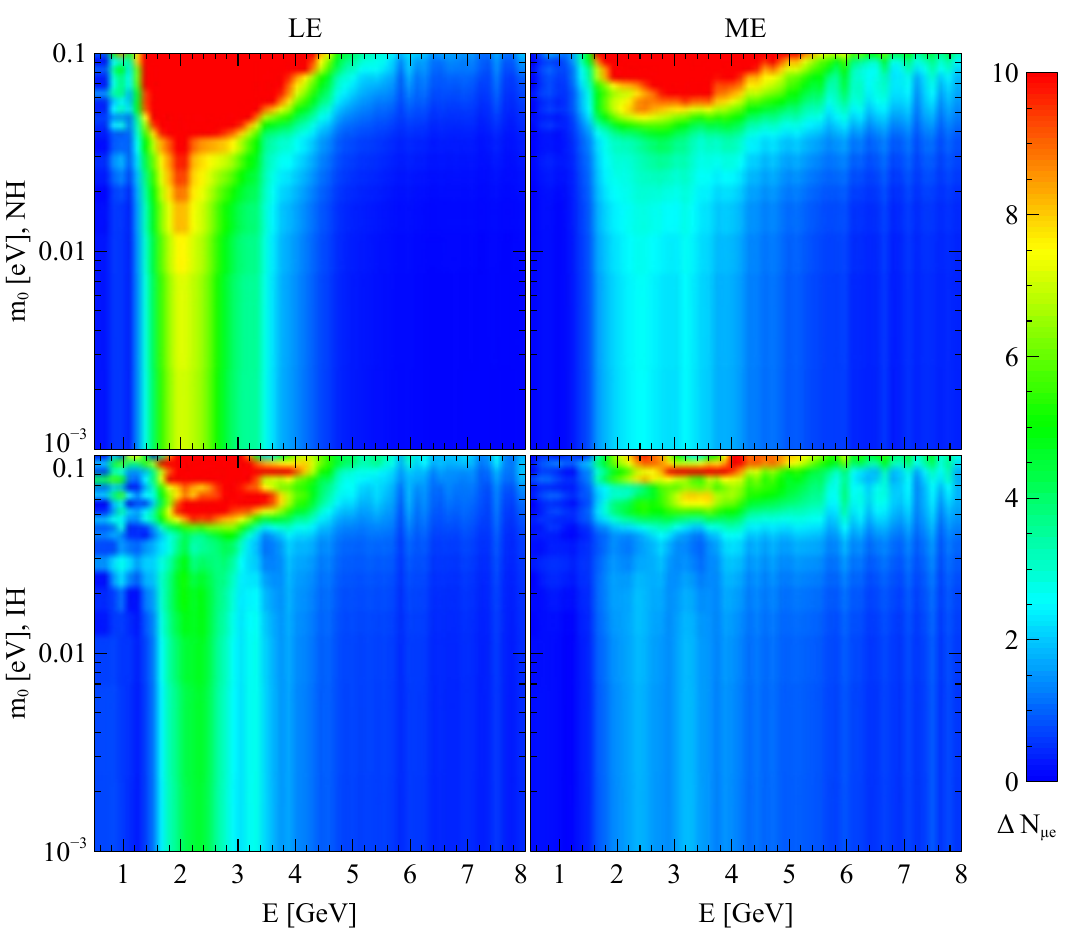}
    \caption{\footnotesize{This shows the heatplot for the absolute difference of $\nu_{e}$-like events for the standard and LED case:  $\Delta N_{\mu e} = |N_{\mu e}^{\text{std}} - N_{\mu e}^{\text{LED}}|$ in the plane of $m_{0}$ and E. 
    The left (right) column corresponds to the use of LE (ME) beam for the full 13 yrs. of runtime, while the top (bottom) row indicates NH (IH). 
   The best-fit values of the standard oscillation parameters (Table \ref{tab:parameters}) were used in generating this figure. 
    }}
    \label{fig:event_heatplot_mue}
\end{figure}
\begin{figure}
    \centering
    \includegraphics[scale=0.7]{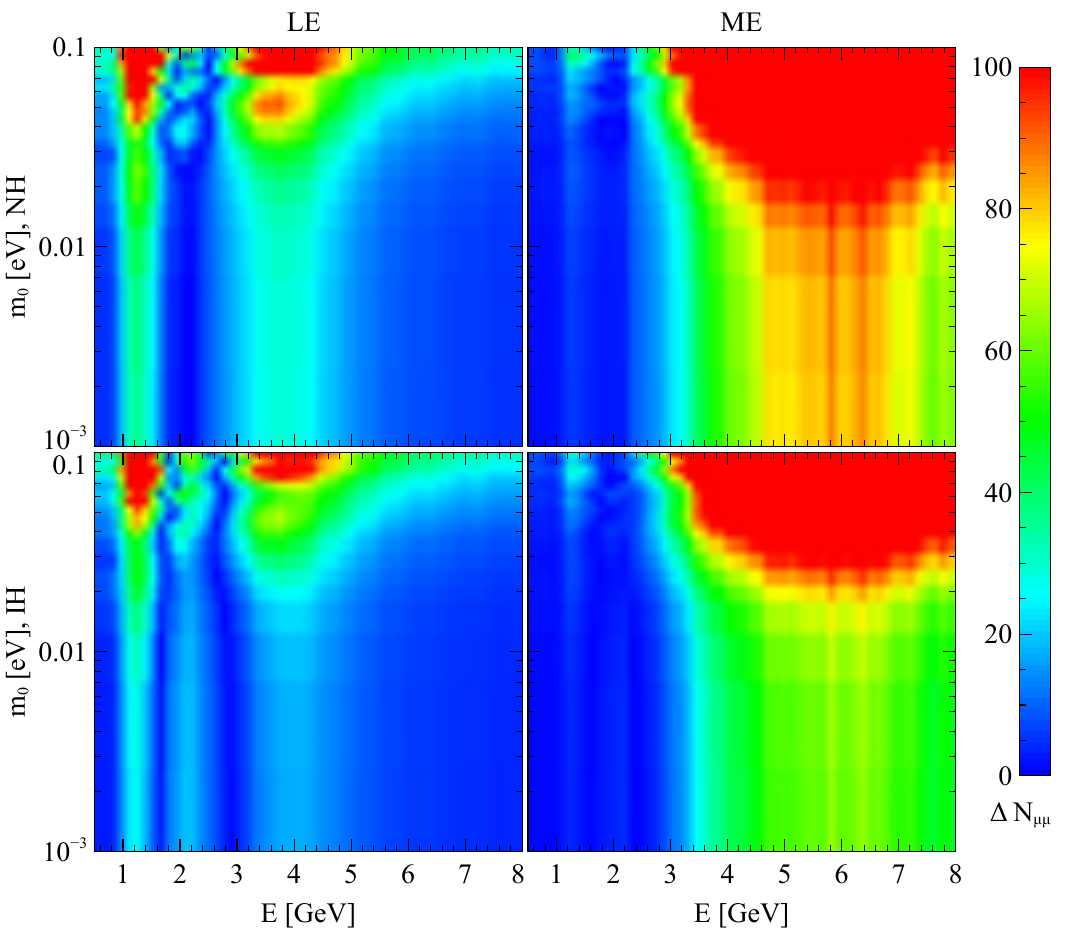}
    \caption{\footnotesize{
    Similar to Fig.\ \ref{fig:event_heatplot_mue} but for the $\nu_{\mu}$-like events.
    }}
    \label{fig:event_heatplot_mumu}
\end{figure}
\begin{figure}
    \centering
    \includegraphics[scale=0.7]{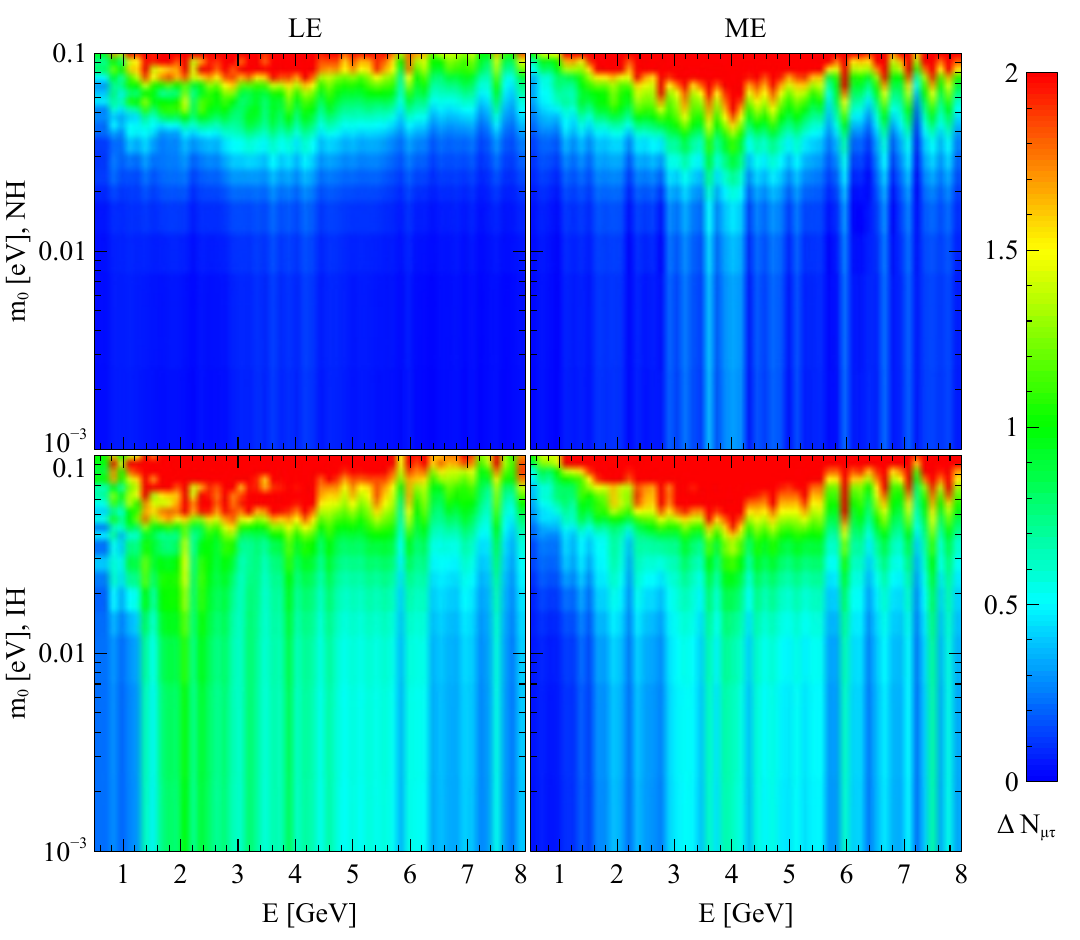}
    \caption{\footnotesize{
    Similar to Figs.\ \ref{fig:event_heatplot_mue} and \ref{fig:event_heatplot_mumu} but for the $\nu_{\tau}$-like events.
    }}
    \label{fig:event_heatplot_mutau}
\end{figure}

To analyse the role of ME more in probing the physics of LED, we now define the following quantification of 
difference of event spectra, 
\begin{equation}
\Delta N_{\alpha\beta} = \big| N_{\alpha\beta}^{\text{std}} - N_{\alpha\beta}^{\text{LED}} \big|,
\label{eq:delta_N}
\end{equation}
where the first and second term in the right hand side denote the number of events for the 
standard and LED case respectively. 
Clearly the quantity $\Delta N_{\alpha\beta}$ is a function of energy. 
We estimate $\Delta N_{\alpha\beta}$ 
for all the energy bins upto 8 GeV at the DUNE FD using either the LE beam or the ME beam alone and generate a heatplot for 
$\Delta N_{\alpha\beta}$ by varying the lightest neutrino mass $m_{0}$ in the range 
$[0.001, 0.1]$ eV. 
 The LED radius $\rad$ is kept fixed at 0.5 $\mu$m throughout.
 Such heatplots are shown in Figs.\ \ref{fig:event_heatplot_mue}, \ref{fig:event_heatplot_mumu} and \ref{fig:event_heatplot_mutau} for the $\nue, \numu$ 
 and $\nutau$ events respectively.  
Clearly, when using LE beam, the deviation of events from the std.\ case in presence of LED is maximum around 
$1.5 - 4$ GeV approximately, where the deviation $\Delta N_{\mu e}$ is quantified to be around $\mathcal{O}(10)$ for $m_{0} \gtrsim 0.04$ eV (top-left panel of Fig.\ \ref{fig:event_heatplot_mue}). 
At higher energies ($\gtrsim 6$ GeV) $\Delta N_{\mu e}$ reduces to very small numbers close to zero. 
Using ME,  the deviation $\Delta N_{\mu e}$ shows a slightly smaller magnitude 
at lower energies, while maintaining some non-zero value (albeit small and $\sim \mathcal{O}(1)$) at $5-8$ GeV. 
Qualitatively similar observations can also be inferred for the IH case (bottom row of Fig.\ \ref{fig:event_heatplot_mue}). 

For the $\mumu$ channel, using the ME beam shows a clear significant advantage 
over the LE beam. 
This is evident from Fig.\ \ref{fig:event_heatplot_mumu} for a significantly wide range 
of energies (approximately 3.5-8 GeV) where the deviation $\Delta N_{\mu\mu}$ can 
attain high values unlike the case where only LE was used. 
 We note that for a large number 
of energy bins (with $4.5 \lesssim E \lesssim 7$ GeV), the use of ME beam can easily give $\Delta N_{\mu\mu} \sim \mathcal{O}(100)$ for even very small values of $m_{0}$ upto 0.001 eV (top right panel of Fig.\ \ref{fig:event_heatplot_mumu}). 
In contrast, with the LE beam $\Delta N_{\mu\mu}$ is much smaller for most of the energy bins and can  reach $\mathcal{O}(100)$ only for very high $m_{0}$ ($\gtrsim 0.07$ eV) and 
only at a relatively narrow energy windows around 4 GeV and 1.5 GeV. 

Due to large backgrounds, the efficiencies for detecting $\nutau$ events at DUNE FD are 
small. 
The capability to differentiate between the std. and LED events are also not very good 
for the $\nutau$ events. 
Hence $\Delta N_{\mu\tau}$ can reach only upto $\mathcal{O}(1)$ at best, as observed in 
Fig.\ \ref{fig:event_heatplot_mutau}. 
This can be obtained only for high values of $m_{0}$ ($\gtrsim 0.06$ eV), and the ME 
beam may offer only a slight advantage.
\section{Analysis methodology}
\label{sec:chisq}
We explore DUNE's capability to probe the LED parameter space by performing a
 $\chisq$ analysis. 
 We use GLoBES~\cite{Huber:2004ka, Huber:2007ji} for the $\chisq$ analysis, 
using the latest GLoBES configuration files corresponding to the Technical Design Report (TDR) by the DUNE collaboration~\cite{DUNE:2021cuw}. 
This analysis was performed at DUNE FD, after taking into account the rate-based constraints at ND 
(without explicitly simulating the ND)~\cite{DUNE:2020fgq}. 
Electron neutrino appearance signals (CC), muon neutrino disappearance signals (CC), as well as neutral current (NC) backgrounds and tau neutrino appearance backgrounds (along with the corresponding systematics/efficiencies etc.) are already included in the configuration files.
In the present analysis, in addition to events from $\mue$ and $\mumu$ channels, we consider the $\mutau$ 
channel. 
Charged current interaction of an incoming $\nu_{\tau}$ produces a $\tau$ lepton, which decays 
hadronically (with a branching fraction $\sim 65\%$) or leptonically (with a branching fraction of $\sim 35\%$). 
The analysis of the hadronic decay channel involves the capability of the detector to study the 
resulting pions and kaons as well separating NC neutrino background.
Following  \cite{deGouvea:2019ozk}, we have used an efficiency to identify $30\%$ hadronically decaying $\tau$ events. 
On the other hand, the leptonic decay channels of $\tau$ ($\tau^{-} \to e^{-}\bar{\nu}_{e}\nu_{\tau}$; $\tau^{-} \to \mu^{-}\bar{\nu}_{\mu}\nu_{\tau}$) are more difficult to analyse, due to the large background mainly consisting of $\nu_{e}$-CC and $\nu_{\mu}$-CC respectively (along with backgrounds from NC and contaminations due to wrong sign leptons.).
Following \cite{Ghoshal:2019pab}, we have taken $15\%$ efficiency to detect $\nutau$ events for leptonically decaying $\tau$ lepton.
We acknowledge that our implementation of the $\nu_{\tau}$ channel as a signal is conservative in nature. Nevertheless, this provides a small but non-negligible statistics in terms of events and $\chisq$ sensitivity (see Sec.\ \ref{sec:results}). 
Using a much more sophisticated analysis of $\nu_{\tau}$ appearance channel at DUNE (for instance, by implementing jet-
clustering algorithms and machine learning techniques, as has been discussed in \cite{Machado:2020yxl}), one certainly expects to exploit the capabilities of $\nutau$ appearance channel further. 
$\nu_{\tau}$ interaction cross-sections are set to zero in the original GLoBES TDR configuration files. 
We include the non-zero $\nu_{\tau}$ cross-sections from a slightly older GLoBES configuration~\cite{DUNE:2016ymp} and also use new \textit{rules} for $\nu_{\tau}$ channel signals in GLoBES following the discussions above. 

 Below we discuss in detail about the approximate analytical form of the $\chisq$ and its various terms. 
\begin{align}
\label{eq:chisq}
\Delta \chi^{2}(\bar{p}^{\text{fit}})  &= \underset{(p^{\text{fit}}-\bar{p}^{\text{fit}}; \eta)}{{\text{min}}} 
\bigg[
\color{black}
\underbrace{\color{black}
2\sum_{x}^{\text{mode}}\sum_{j}^{\text{channel}}\sum_{i}^{\text{bin}}
\bigg\{
N_{ijxy}^{\text{LED}}(p^{\text{fit}};\eta) - N_{ijxy}^{\text{std}}(p^{\text{data}}) 
+ N_{ijxy}^{\text{std}}(p^{\text{data}}) \ln\frac{N_{ijxy}^{\text{std}}(p^{\text{data}})}{N_{ijxy}^{\text{LED}}(p^{\text{fit}};\eta)} 
}_{\text{statistical}}
\bigg\} 
\color{black}
\nonumber \\
&+ 
\color{black}
\underbrace{\color{black}
\sum_{l}\frac{(p^{\text{data}}_{l}-p^{\text{fit}}_{l})^{2}}{\sigma_{p_{l}}^{2}}
}_{\text{prior}}
\color{black}
+ 
\color{black}
\underbrace{\color{black}
\sum_{k}\frac{\eta_{k}^{2}}{\sigma_{k}^{2}}
}_{\text{systematics}}
\color{black}
\bigg],
\end{align}
where 
the index $i$ is summed over the energy bins in the range $0.5-18$ GeV\,\footnote{We have a total of $66$ energy bins in the range $0.5-18$ GeV: $60$ bins each having a width of 
$0.125$ GeV in the energy range of $0.5 - 8$ GeV and $6$ bins
with variable widths beyond $8$ GeV~\cite{DUNE:2021cuw}.}. 
The index $j$ corresponds to three oscillation channels ($\mue, \mumu, \mutau$) while the index $x$ runs over the modes ($\nu$ and $\bar{\nu}$).
$N^{\text{std}}$ (treated as \textit{data}) and $N^{\text{LED}}$ (treated as \textit{fit}) are the set of events corresponding to the standard and LED cases respectively. 
The terms in the first row of the right-hand-side of Eq.~\eqref{eq:chisq} correspond to the statistical contribution. 
The first two terms in the statistical contribution to the $\chisq$ constitute the algebraic difference 
($N^{\text{LED}} - N^{\text{std}}$) while the last term (\ie, the logarithmic-term) corresponds to the fractional difference between the two sets of events%
\footnote{
Note that the definition of $\chisq$ described in Eq.\ \ref{eq:chisq} is Poissonian in nature. In the limit of large events,  this reduces to the Gaussian form : 
\begin{align*}
\lim_{N \to \infty} \Delta \chi^{2}(\bar{p}^{\text{fit}})  \simeq \underset{(p^{\text{fit}}-\bar{p}^{\text{fit}}; \eta)}{{\text{min}}} 
\Bigg[\sum_{x}^{\text{mode}}\sum_{j}^{\text{channel}}\sum_{i}^{\text{bin}}
\frac{
\Big(
N_{ijxy}^{\text{LED}}(p^{\text{fit}};\eta) - N_{ijxy}^{\text{std}}(p^{\text{data}})  
\Big)^{2}}{N_{ijxy}^{\text{std}}(p^{\text{data}})}
+ \text{prior} + \text{systematics}\Bigg].
\end{align*}}. %
Note that $p^{\text{data}}$ ($\{\ta, \tb, \tc, \da, \sdm, \ldm \}$) and $p^{\text{fit}}$ ($\{\ta, \tb, \tc, \da, \sdm, \ldm, m_{0}, \rad \}$) refer to the set of oscillation parameters for the calculation of $N^{\text{std}}$ and $N^{\text{LED}}$ respectively. 
\textit{Data} (\ie, $N^{\text{std}}$) is generated for the standard three-neutrino case by considering the best-fit values of the oscillation parameters from Table~\ref{tab:parameters}.  
The events in the LED case ($N^{\text{LED}}$) are then fit to the \textit{data} by varying 
the set of LED parameters $\bar{p}^{\text{fit}} : \{m_{0}, \rad \}$. 
The set of standard oscillation parameters ($p^{\text{fit}} - \bar{p}^{\text{fit}}$) are also varied 
in the \textit{fit} (using the uncertainties listed in Table \ref{tab:parameters}).

The two terms in the second line of Eq.\ \ref{eq:chisq} correspond to the prior and systematics respectively. 
The \textit{prior} term accounts for the penalty of the $l$ 
number of \textit{fit} parameters deviating away from the 
corresponding $p^{\text{data}}$. 
The degree of this deviation is controlled by  $\sigma_{pl}$ which is the uncertainty in 
the prior measurement of the best-fit values of $p^{\text{data}}$ (see the last column of Table \ref{tab:parameters} 
for the values used in the present analysis). 
The  \textit{systematics}-term accounts for the variation of the systematic/nuisance parameters.
$\eta$ is the set of values of $k$-systematics parameters $\{\eta_{1}$, $\eta_{2}$, $\dots$ $\eta_{k}\}$  while 
$\sigma_k$ is the uncertainty in the corresponding  systematics.

This way of treating the nuisance parameters in the $\chisq$ calculation is known as the {\it{method of pulls}}~\cite{Huber:2002mx,Fogli:2002pt,GonzalezGarcia:2004wg,Gandhi:2007td}. 
Regarding the systematics~\cite{DUNE:2021cuw}, the $\nu_{e}$ and $\bar{\nu}_{e}$ signal modes have a normalization uncertainty of $2\%$ each, whereas the $\nu_{\mu}$ and $\bar{\nu}_{\mu}$ signals have a normalization uncertainty of $5\%$ each. The $\nu_{\tau}$ and $\bar{\nu}_{\tau}$ signals have a normalization uncertainty of $20\%$ each. 
The background normalization uncertainties lie within $5\%-20\%$ and include correlations among various sources of background (coming from beam $\nu_{e}/\bar{\nu}_{e}$ contamination, flavour misidentification, NC and $\nu_{\tau}$).
The final estimate of the $\chisq$ as a function of the set of desired parameters ($\bar{p}^{\text{fit}}$) (\ie, as a function of the LED parameters $m_{0}$ and $\rad$) for a given set of fixed parameters $p^{\text{data}}$ is obtained 
after minimizing the entire quantity within the square bracket in Eq.\ \ref{eq:chisq} over the relevant set of rest of the \textit{fit} parameters $p^{\text{fit}}-\bar{p}^{\text{fit}}$ (\ie, over all the standard oscillation parameters), as well as over the systematics $\eta$. 
This minimization is also referred to as marginalization over the set $\{p^{\text{fit}}-\bar{p}^{\text{fit}}; \eta\}$.  
Technically, this  procedure is the frequentist method of hypotheses testing~\cite{Fogli:2002pt, Qian:2012zn}.  
\section{Results}
\label{sec:results}
\begin{figure}[htb]
    \centering
    \includegraphics[scale=0.75]{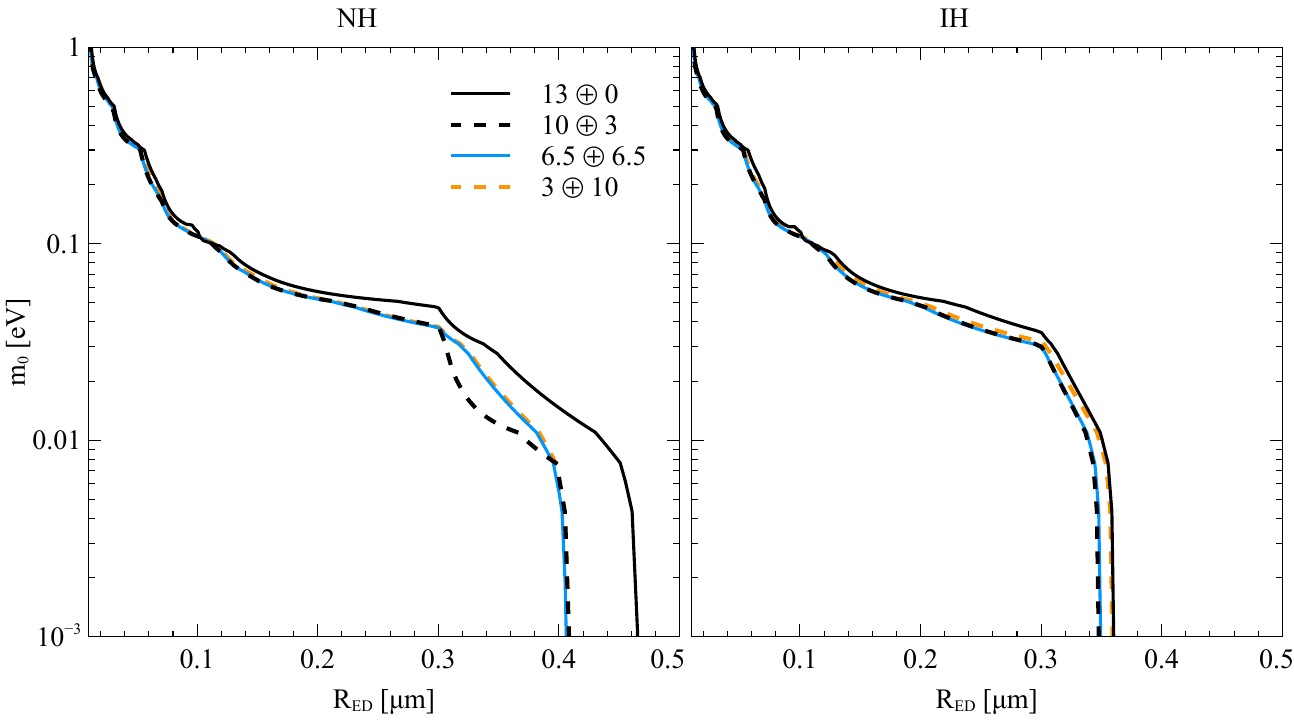}
    \caption{\footnotesize{This shows $\chisq$ contours at $90\%$ confidence level (C.L.)  obtained at DUNE FD at  in the parameter space of $m_{0}-\rad$ for various combinations of LE and ME beam. The legends show the runtime combinations (in yrs.) used for LE and ME beams. For \eg, the black solid curve corresponds to 13 yrs. of runtime in LE only; the black 
    dashed curve refers to 10 yrs. of runtime in LE beam combined with 3 yrs. of runtime 
    using ME beam. Runtime in each beam are equally distributed into $\nu$ and $\bar{\nu}$ 
    modes. The left (right) columns indicates the case of NH (IH).
    }}
    \label{fig:chisq_optimization}
\end{figure}
From the discussions in Sec.\ \ref{sec:event}, we anticipate that at higher energies, ME beam should 
give more sensitivity to LED parameters, while the sensitivities at lower 
energy are expected to be better using the LE beam. 
In order to properly utilize the entire energy range, we seek to optimize the combinations of runtime using LE and ME beams together. 
In Fig.\ \ref{fig:chisq_optimization}, we show the $90\%$ C.L. $\chisq$ contours (\ie, $\chisq = 4.61$ at two degrees of freedom) in the parameter space of $m_{0}-\rad$ at DUNE FD for different combinations of runtime (in yrs.) using LE and ME beam together such that the total runtime remains fixed at 13 yrs.
We marginalise over all six standard oscillation parameters $\ta, \tb, \tc, \da, \sdm, \ldm$ by varying the 
value of $\tc$ to include both the octants and assuming the neutrino mass hierarchy to be either NH or IH 
at a time. We take the prior uncertainties of the oscillation parameters as tabulated in Table \ref{tab:parameters} with $\da$ allowed to vary over the entire range of $[-\pi, \pi]$.  
From Fig.\ \ref{fig:chisq_optimization}, we note that the use of ME beam even for a small runtime 
increases the sensitivity to LED parameters in comparison to LE beam alone, especially for the NH scenario (left panel of Fig.\ \ref{fig:chisq_optimization}). 
For \eg , for $m_{0} \simeq 0.001$ eV, DUNE can exclude $\rad \gtrsim 0.47$ $\mu$m with LE beam alone. 
But using a combination of LE + ME beam, the exclusion capability becomes stronger by almost $13\%$ ($\rad \gtrsim 0.41$ $\mu$m). 
The role of the ME beam is most significant in the region $m_{0} \lesssim 0.01$ eV. 
For values of $m_{0}$ ($\gtrsim 0.1$ eV), all the combinations (including the LE beam alone) give similar sensitivities
For the IH scenario, the sensitivity with LE alone is already better ($\rad \gtrsim 0.36$ $\mu$m at $m_{0} \simeq 0.001$ eV) than the NH case. 
Sharing more than half of the total runtime with the ME beam offers a slight improvement of 
the bounds (by a factor of $5-6\%$ roughly).   
But a smaller runtime with the ME beam does not practically offer improvement compared to running the experiment with LE beam alone (orange dashed curve in the right panel of Fig.\ \ref{fig:chisq_optimization}). 
Similar to the NH case, the role of the ME beam is not relevant above $m_{0} \simeq 0.1$ eV.
From Fig.\ \ref{fig:chisq_optimization}, it appears that a runtime combination of $(10 \oplus 3)$ 
yrs. in the (LE $\oplus$ ME) beam combination offers the best optimized combination to probe the LED parameter space for both NH 
and IH scenario. 
It needs to be mentioned that all the sensitivities to $\rad$ are approximately independent 
of $m_{0}$ for $m_{0} \lesssim 0.01$ eV.
Below we compare the results of this optimized combination in more detail.

\begin{figure}[htb]
    \centering
    \includegraphics[scale=0.75]{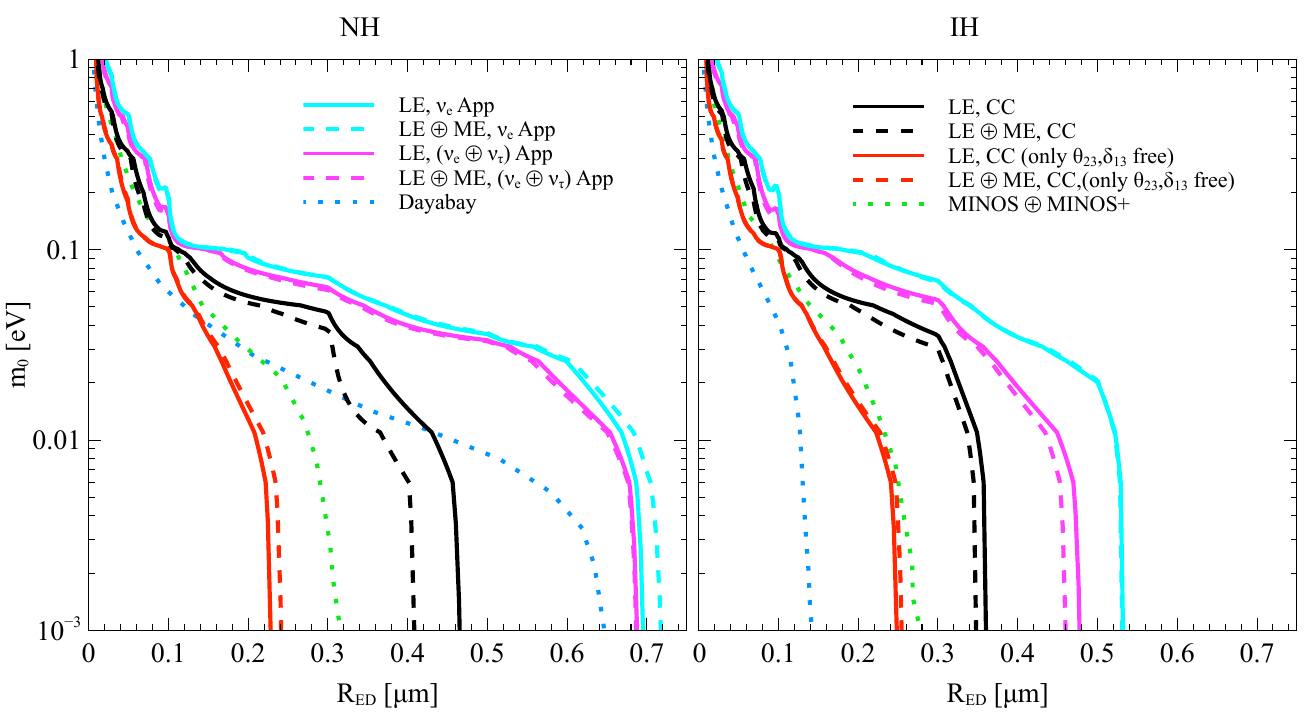}
    \caption{\footnotesize{This shows $90\%$ C.L. $\chisq$ contours obtained with DUNE FD in the parameter space of $m_{0}-\rad$. The solid curves show the results with LE beam alone (13 yrs. of runtime), while the dashed curves show the results when the optimized runtime combination of 10 yrs. with LE beam was used in conjunction with 3 yrs. with ME beam. The cyan and magenta curves depict the case when $\nue$ appearance channel and $(\nue + \nutau)$ appearance channel was used respectively for the analysis. The black curves 
    show the results when all three CC channels ($\nue$ appearance, $\nutau$ appearance and $\numu$ disappearance) were used together. For the cyan, magenta and black curves, 
    all six standard oscillation parameters were considered free and marginalized over. 
    The red curves depict the results when only two standard oscillation parameters $\tc$ and $\da$ were marginalized while the rest were kept fixed at their bestfit values (Table \ref{tab:parameters}). The left (right) panel corresponds to the NH (IH) case. 
    The dotted contours refer to the $90\%$ exclusion contours using the  
     dataset of MINOS/MINOS+ (green) and Daya Bay (blue) respectively (as estimated in ~\cite{Forero:2022skg}). 
    }}
    \label{fig:chisq}
\end{figure}
Fig.\ \ref{fig:chisq} shows the $90\%$ C.L. $\chisq$ contours in the LED parameter space 
 at DUNE FD. 
It compares the sensitivities obtained from LE beam alone (13 yrs.\ of runtime) with that 
obtained from the optimized runtime combination of $(10 \oplus 3)$ yrs.\ using (LE $\oplus$ ME) beam together, by analyzing the sensitivities contributed by various oscillation channels. 
At first we perform a conservative analysis by keeping all six standard oscillation parameters free and marginalizing over all of them (cyan, magenta, black contours in Fig.\ \ref{fig:chisq}).
In the following we discuss various noteworthy observations, by considering the NH scenario first (left panel Fig.\ \ref{fig:chisq}). 
The $\nue$ appearance channel alone can constrain $\rad$ upto roughly 0.69 $\mu$m for $m_{0} = 0.001$ eV when using LE beam only for the NH scenario. 
Combining this channel with the ME beam actually worsens the sensitivity slightly. 
Considering $\nutau$ appearance channel into the analysis proves to be of very small impact as well (whether using LE beam alone or the optimized combination of LE $\oplus$ ME beam). 
Adding the $\numu$ disappearance channel using the LE beam alone significantly strengthens the constraints by excluding
 $\rad \gtrsim 0.47$ $\mu$m, an improvement of roughly $33\%$ over what can be achieved using 
$\nue$ appearance channel using LE beam only. 
We emphasize that this constraint obtained using the latest DUNE TDR configurations and three oscillation channels is slightly better when compared to other projected DUNE constraints~\cite{Berryman2016, DUNE:2020fgq, Roy:2023dyq}.  
The optimized combination of runtime using (LE $\oplus$ ME) beams tightens the 
constraints further by a factor of $13\%$ to a value $\rad \gtrsim 0.41$ $\mu$m at $m_{0} = 0.001$ eV
(as already mentioned in the context of Fig.\ \ref{fig:chisq_optimization}.). 
Use of $\mumu$ oscillation channel at DUNE begins to offer better sensitivity below $m_{0} \lesssim 0.01$ eV than the Daya Bay data.

We also perform the analysis with an optimistic case, - assuming that all the standard oscillation 
parameters except $\tc$ and $\da$ will be known to a very high accuracy when the actual data taken by DUNE will be analyzed. 
In the optimistic scenario we only marginalize over $\tc$ and $\da$ and keep the rest of the 
oscillation parameters fixed to their best fit values in the \textit{fit}. 
We show the consequent sensitivities (taking all three oscillation channels together) as the red lines in Fig.\ \ref{fig:chisq}. 
Using LE beam alone, such an analysis puts a strong constraint and excludes $\rad \gtrsim 0.23$ $\mu$m at $m_{0} \simeq 0.001$ eV (an improvement of almost $50\%$ over the corresponding conservative assumption). 
The optimized combination of (LE $\oplus$ ME) does not prove to be very beneficial in the 
optimistic assumption, - rather worsening the exclusion region slightly: $\rad \gtrsim 0.24$ $\mu$m at $m_{0} \simeq 0.001$ eV. 
We should mention that in a very recent analysis of LED parameters using simulated data at DUNE FD with 
a high energy beam alone where the authors have marginalised over $\da, \ldm, \tc$ with the octant of $\tc$ fixed, gives a similar constraint at $2\sigma$ C.L.~\cite{Giarnetti:2024mdt}.  

Crucially we observe that with the optimistic assumption DUNE can reach into 
the sensitivity regions for MINOS/MINOS+ for almost all values of $m_{0}$ under consideration. 
The future projections of other LBL experiments~\cite{Roy:2023dyq} such as T2HK and ESS$\nu$SB give constraints 
($\rad \gtrsim 0.45$ $\mu$m and $\rad \gtrsim 0.6$ $\mu$m respectively at $m_{0} \simeq 0.001$ eV)
comparable to what the present analysis finds using only LE beam at DUNE.  
The analyses of Gallium experiments GALLEX~\cite{GALLEX:1997lja}, SAGE~\cite{SAGE:1998fvr}, BEST~\cite{Barinov:2021asz} give somewhat weaker bounds on LED and can be well-excluded by DUNE, as well as reactor experiments like Daya Bay. 
Though the data from the direct neutrino mass measurement experiment KATRIN~\cite{KATRIN:2001ttj} only constrains 
LED very weakly, the combined analysis of KATRIN with MINOS/MINOS+ and Daya Bay can give very strong bounds on $\rad$~\cite{Forero:2022skg}, - which are comparable to what can be achieved using the optimistic assumptions at DUNE. 
 Apart from the neutrino sector, the bounds on LED have also been obtained from tabletop gravitational experiments~\cite{Long:1998dk, Krause:1999ry, Fischbach:2001ry, Adelberger:2002ic, Decca:2003td}, collider experiments~\cite{Rizzo:1998fm, Hewett:1998sn, D0:2000cve, DELPHI:2000ztm, DELPHI:2008uka}, and also from astrophysical~\cite{Cullen:1999hc, Barger:1999jf, Hanhart:2000er, Hannestad:2001jv, Hannestad:2003yd, Mohapatra:2003ah, Feng:2003nr, Cacciapaglia:2003dx} and cosmological data~\cite{Hall:1999mk, Hannestad:2001nq, Fairbairn:2001ct, Fairbairn:2001ab, Goh:2001uc}. 
 The tabletop experiments give constraints on $\rad$ which are about two orders of magnitude weaker ($\rad \gtrsim 37$ $\mu$m at $95\%$ C.L.) than 
 what can be achieved at DUNE
 Using astrophysical data, very strong constraints ranging in $\rad \gtrsim 0.16 - 916$ nm have been obtained. 
 However, these limits depend on the technique and some assumptions~\cite{ParticleDataGroup:2022pth} in the analyses.

For the IH scenario (right panel of Fig.\ \ref{fig:chisq}), the sensitivities obtainable by the 
$\nue$ appearance ($\rad \gtrsim 0.53$ $\mu$m) and $(\nue \oplus \nutau)$ appearance channels ($\rad \gtrsim 0.48$ $\mu$m) are better compared 
to the NH case. 
Consequently the bound achieved with all three CC channels are better than the NH case, 
even though the contribution from the dominant channel $\mumu$ is largely insensitive 
of neutrino mass hierarchy~\footnote{For small $m_{0}$ ($\lesssim 0.1$ eV), in the NH case we have $\xi_{3} >> \xi_{1}, \xi_{2}$. 
Thus from Eqs.\ \ref{eq:phi_00} and \ref{eq:phi_0n} it is clearly seen that the LED effects on probability in Eq.\ \ref{eq:prob} 
are dominated by terms when the index $j$ equals 3. 
But such terms are also multiplied by the PMNS matrix terms such as $|U_{\alpha 3}|^{2}$ ($\alpha=e,\mu,\tau$). 
For the $\mue$ channel (as well as in $\bar{\nu}_{e} \to \bar{\nu}_{e}$ channel in reactor neutrino experiments), the very small magnitudes of $|U_{e3}|^{2}$ thus suppresses the dominant LED terms for NH. 
But for IH, $\xi_{3} << \xi_{1}, \xi_{2}$, and the corresponding PMNS matrix factors are $|U_{\alpha 1}|^{2}$ 
and $|U_{\alpha 2}|^{2}$, - which are reasonably large for all $\alpha = e,\mu,\tau$. 
This makes the sensitivity to LED parameters in the $\mue$ channel  stronger in the IH case. 
But due to absence of such suppression by the PMNS matrix terms, the other oscillation channels are mostly independent of neutrino mass hierarchy.}. 
Further, we note that the combination with ME beam offers only a slight improvement of results, especially for $\mutau$ and $\mumu$ channels. 
Thus the optimistic assumptions increase the sensitivities significantly, in the IH case, by almost $30\%$. 
All the constraints obtained on $\rad$ for the lowest $m_{0}$ value ($0.001$ eV) are 
tabulated together in Table \ref{tab:bounds}.
 \begin{table}[h]
\centering
\begin{tabular}{| c | c | c | c | c |}
\hline
&&&& \\
Neutrino  & Marginalization  &  & Beam & Bounds on  \\
Mass & Assumption & Channel & used & $\rad$ [$\mu$m]\\
Hierarchy & & &  & ($90\%$ C.L.)\\
&&&& \\
\hline
\multirow{8}{*}{NH} &  & \multirow{2}{*}{$\mue$} & LE & 0.69 \\ 
\cline{4-5}
&  & & LE $\oplus$ ME & 0.72\\
\cline{3-5}
& All six std. osc. & \multirow{2}{*}{$(\mue) \oplus (\mutau)$} & LE & 0.68 \\
\cline{4-5}
& parameters free &  & LE $\oplus$ ME & 0.68\\
\cline{3-5}
&& \multirow{2}{*}{$(\mue) \oplus (\mutau) \oplus (\mumu)$} & LE & 0.47 \\
\cline{4-5}
&&  & LE $\oplus$ ME & 0.41\\
\cline{2-5}
& \multirow{2}{*}{$\tc, \da$ free} & \multirow{2}{*}{$(\mue) \oplus (\mutau) \oplus (\mumu)$} & LE & 0.23 \\
\cline{4-5}
&&& LE $\oplus$ ME & 0.24 \\
\hline
\multirow{8}{*}{IH} & . & \multirow{2}{*}{$\mue$} & LE & 0.53 \\ 
\cline{4-5}
&  & & LE $\oplus$ ME & 0.53\\
\cline{3-5}
& All six std. osc & \multirow{2}{*}{$(\mue) \oplus (\mutau)$} & LE & 0.48 \\
\cline{4-5}
& parameters free &  & LE $\oplus$ ME & 0.46\\
\cline{3-5}
&& \multirow{2}{*}{$(\mue) \oplus (\mutau) \oplus (\mumu)$} & LE & 0.36 \\
\cline{4-5}
&&  & LE $\oplus$ ME & 0.34\\
\cline{2-5}
& \multirow{2}{*}{$\tc, \da$ free} & \multirow{2}{*}{$(\mue) \oplus (\mutau) \oplus (\mumu)$} & LE & 0.25 \\
\cline{4-5}
&&& LE $\oplus$ ME & 0.26 \\
\hline
\end{tabular}
\caption{\footnotesize{\label{tab:bounds}
The table shows the bounds on $\rad$ [$\mu$m] (as read off from Fig.\ \ref{fig:chisq}) beyond which LED is excluded at $90\%$ C.L. for $m_{0} \simeq 0.001$ eV.
}}
\end{table}

Finally it should be mentioned that explicitly simulating an ND and performing the analysis to a combined 
dataset at FD and ND should boost the LED constraints considerably. But this demands careful consideration 
of systematic uncertainties, especially the bin-by-bin shape related systematics and consequently a considerable amount of computing power. This is beyond the scope 
of the present analysis and we leave it as a future work. 
\section{Summary and discussions}
\label{sec:conclusion}
The LED model was proposed as an elegant solution to the hierarchy problem. It also 
offers a natural explanation of neutrino mass generation. All the SM particles including the 
left handed neutrinos lie on the familiar 4-dimensional spacetime known as brane. 
The brane is embedded within a higher dimensional ($4+n$) spacetime (with $n$ extra 
spatial dimension) known as bulk. 
All the particles that are singlets under the SM symmetries (such as gravitons, right handed neutrinos) can propagate into the bulk. 
We consider effectively 1 extra spatial dimension. 
The couplings of 
the three right handed neutrino fields lying in the $(4+1)$ dimension with the three  
left handed SM neutrino fields in the 4 dimensional spacetime, are suppressed by the large volume of the extra spatial dimension and generates tiny masses for the active neutrinos in a natural way. 
The three five dimensional right handed neutrino fields behave like infinite number of Kaluza-Klein (KK) modes from a four-dimensional point of view after the compactification of the fifth 
dimension. 
The compactification radius $\rad$ and the lightest Dirac neutrino mass ($m_{0}$) 
(in addition to the six standard oscillation parameters) determine the neutrino oscillation 
probability in presence of LED. 

In the present work we study the capabilities of DUNE FD to probe the LED parameters 
$m_{0}$ and $\rad$ by using combination of different beam tunes. 
After a brief overview of the theoretical basics of LED in the context of neutrino oscillation (Sec.\ \ref{sec:theory}), 
 we analyze the oscillation probability for the three channels $\mue, \mumu$ and 
$\mutau$ (Sec.\ \ref{sec:prob}). 
We note the two-fold impact of LED, - namely an overall reduction of the 
magnitude of oscillation probability for all flavours, and introduction of additional frequencies 
(due to transition into the KK modes) in the probability spectrum. 
The impacts of LED seem to be more pronounced at higher energies ($\gtrsim 5$ GeV), 
which is especially prominent for the $\mumu$ channel.
In Sec.\ \ref{sec:event}, we show the two beams used in the present analysis. 
The first beam is the standard low energy (LE) tuned neutrino beam at DUNE peaking at low energies ($2-3$ GeV) and falling sharply beyond 4 GeV. 
The second beam is the $\nutau$-optimized beam that has a broad peak at energies 
($3-5$ GeV) and falling much slowly beyond that. 
We compare the event spectra using LE beam or ME beam at-a-time for the three channels both in standard case and in presence of LED, and highlight how the ME beam offers more statistics at higher energies, - a fact which gives order-of-magnitude higher events for the $\mumu$ channel. 
We then proceed to generate heatplots of the absolute difference of events (generated using either the LE beam or the ME beam) between the standard and LED case in the space of $m_{0}$ and neutrino energy $E$. 
The heatplots give a clear idea about how using ME beam is expected to help provide 
higher difference of events between the standard and LED case at higher energy bins, - thus potentially providing 
more sensitivities for a large range of values of $m_{0}$. 

After a detailed discussion of the $\chisq$ methods (Sec.\ \ref{sec:chisq}) we proceed to 
discuss our main results in Sec.\ \ref{sec:results}. 
At first, we generate the $90\%$ C.L. $\chisq$ contours in the plane of $m_{0}-\rad$ 
by for various combinations of runtimes shared between LE and ME beams in order to 
optimize the runtime combinations. 
We find that rather than using the LE beam alone, using a combination of (LE $\oplus$ ME) 
with $(10+3)$ yrs. of runtime gives better sensitivities to the LED parameters for both NH and IH case. 
We then analyze the contributions of the three oscillation channels to the sensitivities and show how the 
combination of (LE $\oplus$ ME) beams impacts the channels individually. 
As expected, we find the role of $\mumu$ channel to be the most crucial and estimate  
 the bounds on $\rad$ as obtained using LE alone or with a combination of (LE $\oplus$ ME). 
 We also perform an analysis using optimistic assumptions that all the values of the standard oscillation parameters except $\tc, \da$ will be known to a very high precision when the analysis of the actual DUNE data will take place, and find out that the bounds on $\rad$ 
 strengthens significantly for both NH (by $50\%$) and IH (by $30\%$). 
 \appendix
\renewcommand{\theequation}{\thesection.\arabic{equation}}
\setcounter{equation}{0}
\renewcommand{\thesection}{\Alph{section}}
\section{Appendix: Details of implementation of the physics of LED into GLoBES}
\label{appendix_a}
To implement the physics of LED, we follow the prescription of GLoBES 
manual (GLoBES version 3.2.18, chapter 8). 
We copy the relevant parts within the GLoBES source code (from the file \texttt{glb\_probability.c}) to our main C++ file. 
The relevant parts that are copied are the following three functions.
\begin{enumerate}
\item 
\texttt{int (*glb\_set\_oscillation\_parameters\_function)(glb\_params p, void* user\_data)},
\item 
\texttt{int (*glb\_get\_oscillation\_parameters\_function)(glb\_params p, void* user\_data)},
\item
\texttt{int (*glb\_probability\_matrix\_function)(double P[3][3], int cp\_sign, double E, int psteps, const double *lengths, const double *densities, double filter\_sigma, void* user\_data)}.
\end{enumerate}
The first two of these functions are then used to include the additional oscillation 
parameters that appear in presence of LED (namely $m_{0}$ and $\rad$) 
in the GLoBES probability engine. Here we follow the prescription implemented in \texttt{example6.c} of the GLoBES example directory. 
The third function above is used for the actual calculation of oscillation probability 
using Eq.\ \ref{eq:prob} (which is also written below for clarity).
\begin{equation}
P_{\nu_{\alpha} \to \nu_{\beta}}^{\text{LED}} = \bigg|
\sum_{j=1}^{3}\sum_{n=0}^{2}U_{\alpha j}^{*}U_{\beta j}(V_{j}^{n})^{2}
\text{exp}\bigg(-i\frac{(m_{j}^{n})^{2}L}{2E} \bigg)
\bigg|^{2}.
\label{eq:prob_appendix}
\end{equation}
Note that we have considered the KK modes upto $n=2$ since the contributions 
for $n>2$ is negligible. 
Here, $U$ is usual $3 \times 3$ PMNS mixing matrix, $L$ is the neutrino propagation distance (1300 km) and $E$ is the neutrino energy. 
To proceed further, we note that expanding Eq.\ \ref{eq:prob_matter} using the 
matrix form for $M$ in Eq.\ \ref{eq:M} upto $n=2$ KK modes, we can arrive at the following equation 
for neutrino evolution in matter in presence of LED (see Appendix A of \cite{Machado:2011jt} for a detailed calculation).
 \begin{equation}
 i\frac{d}{dt}\begin{pmatrix}
 \nu_{1}^{\prime 0} \\ \nu_{2}^{\prime 0} \\ \nu_{3}^{\prime 0} \\
 \nu_{1}^{\prime 1} \\ \nu_{2}^{\prime 1} \\ \nu_{3}^{\prime 1} \\
 \nu_{1}^{\prime 2} \\ \nu_{2}^{\prime 2} \\ \nu_{3}^{\prime 2}
  \end{pmatrix}
  = 
  \underbrace{
  \frac{1}{2E\rad^{2}}
  \begin{pmatrix}
  \eta_{1} + X_{11} & X_{12} & X_{13} & \kappa_{1} & 0 & 0 & 2\kappa_{1} & 0 & 0\\
  X_{21} & \eta_{2} + X_{22} & X_{23} & 0 & \kappa_{2} & 0 & 0 & 2\kappa_{2} & 0\\
  X_{31} & X_{32} & \eta_{3} + X_{33}  & 0 & 0 & \kappa_{3} & 0 & 0 & 2\kappa_{3}\\
  \kappa_{1} & 0 & 0 & 1 & 0 & 0 & 0 & 0 & 0\\
  0 & \kappa_{2} & 0 & 0 & 1 & 0 & 0 & 0 & 0\\
  0 & 0 & \kappa_{3} & 0 & 0 & 1 & 0 & 0 & 0\\
  2\kappa_{1} & 0 & 0 & 0 & 0 & 0 & 4 & 0 & 0\\
  0 & 2\kappa_{2} & 0 & 0 & 0 & 0 & 0 & 4 & 0\\
  0 & 0 & 2\kappa_{3} & 0 & 0 & 0 & 0 & 0 & 4\\
  \end{pmatrix}
  }_{\mathcal{H}}
  \begin{pmatrix}
   \nu_{1}^{\prime 0} \\ \nu_{2}^{\prime 0} \\ \nu_{3}^{\prime 0} \\
 \nu_{1}^{\prime 1} \\ \nu_{2}^{\prime 1} \\ \nu_{3}^{\prime 1} \\
 \nu_{1}^{\prime 2} \\ \nu_{2}^{\prime 2} \\ \nu_{3}^{\prime 2}
  \end{pmatrix}.
 \label{eq:nu_evolution}
 \end{equation}
 Here, 
 \begin{align}
 &X_{jk} = 2E\rad^{2}Y_{jk} = 2E\rad^{2}\sum_{\alpha=e,\mu,\tau}U_{\alpha j}^{*}U_{\alpha k}(\delta_{\alpha e}V_{\text{CC}}+V_{\text{NC}}), \nonumber \\
&  \kappa_{j} = \sqrt{2}\xi_{j} = \sqrt{2}m_{j}^{D}\rad,  \nonumber \\
 &  \eta_{j} = (N + 1/2)\kappa_{j}^{2}, \text{\quad $(j,k=1,2,3)$},
  \end{align} 
    such that $N$ is the total number of neutrino mass states upto $n=2$ KK modes (\ie, $N=9$). 
 The determination of the matrix V and the mass-squares $(m^{n}_{j})^{2}$ in Eq.\ \ref{eq:prob_appendix} now boils down to the estimation of 
eigenvectors and eigenvalues 
 of the mass-basis hamiltonian $\mathcal{H}$ indicated in Eq.\ \ref{eq:nu_evolution}.
 This eigenvalue problem is numerically solved using the C++ package \texttt{Armadillo}~\cite{Sanderson2016, mca24030070} and estimate the oscillation probability from Eq.\ \ref{eq:prob_appendix}. 
Finally, before returning the probability value from the function \texttt{int (*glb probability matrix function)()}, the GLoBES low-pass filter is applied (by passing the filter value from the \texttt{glb} file through the variable \texttt{double filter\_sigma} in order to 
reduce the fast oscillations that are not observable by the finite energy resolution of the detector. 
As a final step, the three 
 redefined GLoBES functions mentioned above are registered in the main C++ code by using the function,\\
  \texttt{int glbRegisterProbabilityEngine(int n\_parameters, \\  glb\_probability\_matrix\_function prob\_func, \\ glb\_set\_oscillation\_parameters\_function set\_params\_func, \\ glb\_get\_oscillation\_parameters\_function get\_params\_func, \\
void* user\_data)}. 
This way of including the physics of LED (or any new physics) into GLoBES following the steps prescribed in the manual enables us to use the usual GLoBES commands for the numerical calculation of probability, event spectra and $\chisq$.
\section*{Acknowledgments}
The authors acknowledge support from the grant NRF-2022R1A2C1009686. 
This research was supported by the Chung-Ang University Research Scholarship Grants in 2023. 
MM thanks Samiran Roy for helpful discussions regarding LED. 
MM acknowledges the help of Chang-Hyeon Ha in using the cluster at Chung-Ang University for performing some of the numerical calculations presented in the paper. 
This work reflects the views of the authors and not those of the DUNE collaboration.
\bibliographystyle{JHEP}
\bibliography{reference}

\providecommand{\href}[2]{#2}\begingroup\raggedright\begin{thebibliography}{10}

\bibitem{Super-Kamiokande:1998kpq}
{\scshape Super-Kamiokande} collaboration, \emph{{Evidence for oscillation of
  atmospheric neutrinos}},
  \href{https://doi.org/10.1103/PhysRevLett.81.1562}{\emph{Phys. Rev. Lett.}
  {\bfseries 81} (1998) 1562}
  [\href{https://arxiv.org/abs/hep-ex/9807003}{{\ttfamily hep-ex/9807003}}].

\bibitem{SNO:2001kpb}
{\scshape SNO} collaboration, \emph{{Measurement of the rate of $\nu_e+d \to
  p+p+e^-$ interactions produced by $^8$B solar neutrinos at the Sudbury
  Neutrino Observatory}},
  \href{https://doi.org/10.1103/PhysRevLett.87.071301}{\emph{Phys. Rev. Lett.}
  {\bfseries 87} (2001) 071301}
  [\href{https://arxiv.org/abs/nucl-ex/0106015}{{\ttfamily nucl-ex/0106015}}].

\bibitem{10.5281/zenodo.4726908}
{P. F. De Salas}, {D. V. Forero}, {S. Gariazzo}, {P. Mart{\'\i}nez-Mirav{\'e}},
  {O. Mena}, {C. A. Ternes} et~al., ``{Chi2 profiles from Valencia neutrino
  global fit}.'' \url{http://globalfit.astroparticles.es/}, 2021.
\newblock 10.5281/zenodo.4726908.

\bibitem{deSalas:2020pgw}
P.F.~de~Salas, D.V.~Forero, S.~Gariazzo, P.~Mart{\'\i}nez-Mirav{\'e}, O.~Mena,
  C.A.~Ternes et~al., \emph{{2020 global reassessment of the neutrino
  oscillation picture}},
  \href{https://doi.org/10.1007/JHEP02(2021)071}{\emph{JHEP} {\bfseries 02}
  (2021) 071} [\href{https://arxiv.org/abs/2006.11237}{{\ttfamily
  2006.11237}}].

\bibitem{nufit_globalfit}
{I. Esteban}, {M. C. Gonzalez-Garcia}, {M. Maltoni}, {I. Martinez-Soler}, {J.
  P. Pinheiro} and {T. Schwetz}, ``{NuFIT 6.0}.'' \url{www.nu-fit.org}, 2024.

\bibitem{Esteban:2024eli}
I.~Esteban, M.C.~Gonzalez-Garcia, M.~Maltoni, I.~Martinez-Soler,
  J.a.P.~Pinheiro and T.~Schwetz, \emph{{NuFit-6.0: Updated global analysis of
  three-flavor neutrino oscillations}},
  \href{https://arxiv.org/abs/2410.05380}{{\ttfamily 2410.05380}}.

\bibitem{Capozzi:2018ubv}
F.~Capozzi, E.~Lisi, A.~Marrone and A.~Palazzo, \emph{{Current unknowns in the
  three neutrino framework}},
  \href{https://doi.org/10.1016/j.ppnp.2018.05.005}{\emph{Prog. Part. Nucl.
  Phys.} {\bfseries 102} (2018) 48}
  [\href{https://arxiv.org/abs/1804.09678}{{\ttfamily 1804.09678}}].

\bibitem{Minkowski:1977sc}
P.~Minkowski, \emph{{$\mu \to e\gamma$ at a Rate of One Out of $10^{9}$ Muon
  Decays?}}, \href{https://doi.org/10.1016/0370-2693(77)90435-X}{\emph{Phys.
  Lett. B} {\bfseries 67} (1977) 421}.

\bibitem{Yanagida:1979as}
T.~Yanagida, \emph{{Horizontal gauge symmetry and masses of neutrinos}},
  {\emph{Conf. Proc. C} {\bfseries 7902131} (1979) 95}.

\bibitem{Mohapatra:1979ia}
R.N.~Mohapatra and G.~Senjanovic, \emph{{Neutrino Mass and Spontaneous Parity
  Nonconservation}},
  \href{https://doi.org/10.1103/PhysRevLett.44.912}{\emph{Phys. Rev. Lett.}
  {\bfseries 44} (1980) 912}.

\bibitem{Schechter:1980gr}
J.~Schechter and J.W.F.~Valle, \emph{{Neutrino Masses in SU(2) x U(1)
  Theories}}, \href{https://doi.org/10.1103/PhysRevD.22.2227}{\emph{Phys. Rev.
  D} {\bfseries 22} (1980) 2227}.

\bibitem{Arkani-Hamed:1998jmv}
N.~Arkani-Hamed, S.~Dimopoulos and G.R.~Dvali, \emph{{The Hierarchy problem and
  new dimensions at a millimeter}},
  \href{https://doi.org/10.1016/S0370-2693(98)00466-3}{\emph{Phys. Lett. B}
  {\bfseries 429} (1998) 263}
  [\href{https://arxiv.org/abs/hep-ph/9803315}{{\ttfamily hep-ph/9803315}}].

\bibitem{Antoniadis:1998ig}
I.~Antoniadis, N.~Arkani-Hamed, S.~Dimopoulos and G.R.~Dvali, \emph{{New
  dimensions at a millimeter to a Fermi and superstrings at a TeV}},
  \href{https://doi.org/10.1016/S0370-2693(98)00860-0}{\emph{Phys. Lett. B}
  {\bfseries 436} (1998) 257}
  [\href{https://arxiv.org/abs/hep-ph/9804398}{{\ttfamily hep-ph/9804398}}].

\bibitem{Arkani-Hamed:1998sfv}
N.~Arkani-Hamed, S.~Dimopoulos and G.R.~Dvali, \emph{{Phenomenology,
  astrophysics and cosmology of theories with submillimeter dimensions and TeV
  scale quantum gravity}},
  \href{https://doi.org/10.1103/PhysRevD.59.086004}{\emph{Phys. Rev. D}
  {\bfseries 59} (1999) 086004}
  [\href{https://arxiv.org/abs/hep-ph/9807344}{{\ttfamily hep-ph/9807344}}].

\bibitem{Arkani-Hamed:1998wuz}
N.~Arkani-Hamed, S.~Dimopoulos, G.R.~Dvali and J.~March-Russell,
  \emph{{Neutrino masses from large extra dimensions}},
  \href{https://doi.org/10.1103/PhysRevD.65.024032}{\emph{Phys. Rev. D}
  {\bfseries 65} (2001) 024032}
  [\href{https://arxiv.org/abs/hep-ph/9811448}{{\ttfamily hep-ph/9811448}}].

\bibitem{Dienes:1998sb}
K.R.~Dienes, E.~Dudas and T.~Gherghetta, \emph{{Neutrino oscillations without
  neutrino masses or heavy mass scales: A Higher dimensional seesaw
  mechanism}}, \href{https://doi.org/10.1016/S0550-3213(99)00377-6}{\emph{Nucl.
  Phys. B} {\bfseries 557} (1999) 25}
  [\href{https://arxiv.org/abs/hep-ph/9811428}{{\ttfamily hep-ph/9811428}}].

\bibitem{Dvali:1999cn}
G.R.~Dvali and A.Y.~Smirnov, \emph{{Probing large extra dimensions with
  neutrinos}}, \href{https://doi.org/10.1016/S0550-3213(99)00574-X}{\emph{Nucl.
  Phys. B} {\bfseries 563} (1999) 63}
  [\href{https://arxiv.org/abs/hep-ph/9904211}{{\ttfamily hep-ph/9904211}}].

\bibitem{Barbieri:2000mg}
R.~Barbieri, P.~Creminelli and A.~Strumia, \emph{{Neutrino oscillations from
  large extra dimensions}},
  \href{https://doi.org/10.1016/S0550-3213(00)00348-5}{\emph{Nucl. Phys. B}
  {\bfseries 585} (2000) 28}
  [\href{https://arxiv.org/abs/hep-ph/0002199}{{\ttfamily hep-ph/0002199}}].

\bibitem{Nortier:2020lbs}
F.~Nortier, \emph{{Large star/rose extra dimension with small leaves/petals}},
  \href{https://doi.org/10.1142/S0217751X20501821}{\emph{Int. J. Mod. Phys. A}
  {\bfseries 35} (2020) 2050182}
  [\href{https://arxiv.org/abs/2001.07102}{{\ttfamily 2001.07102}}].

\bibitem{Acciarri:2015uup}
{\scshape DUNE} collaboration, \emph{{Long-Baseline Neutrino Facility (LBNF)
  and Deep Underground Neutrino Experiment (DUNE) Conceptual Design Report
  Volume 2: The Physics Program for DUNE at LBNF}},
  \href{https://arxiv.org/abs/1512.06148}{{\ttfamily 1512.06148}}.

\bibitem{DUNE:2016rla}
{\scshape DUNE} collaboration, \emph{{Long-Baseline Neutrino Facility (LBNF)
  and Deep Underground Neutrino Experiment (DUNE)}: {Conceptual Design Report,
  Volume 4 The DUNE Detectors at LBNF}},
  \href{https://arxiv.org/abs/1601.02984}{{\ttfamily 1601.02984}}.

\bibitem{Hyper-KamiokandeProto-:2015xww}
{\scshape Hyper-Kamiokande Proto-} collaboration, \emph{{Physics potential of a
  long-baseline neutrino oscillation experiment using a J-PARC neutrino beam
  and Hyper-Kamiokande}},
  \href{https://doi.org/10.1093/ptep/ptv061}{\emph{PTEP} {\bfseries 2015}
  (2015) 053C02} [\href{https://arxiv.org/abs/1502.05199}{{\ttfamily
  1502.05199}}].

\bibitem{Hyper-Kamiokande:2016srs}
{\scshape Hyper-Kamiokande} collaboration, \emph{{Physics potentials with the
  second Hyper-Kamiokande detector in Korea}},
  \href{https://doi.org/10.1093/ptep/pty044}{\emph{PTEP} {\bfseries 2018}
  (2018) 063C01} [\href{https://arxiv.org/abs/1611.06118}{{\ttfamily
  1611.06118}}].

\bibitem{ESSnuSB:2013dql}
{\scshape ESSnuSB} collaboration, \emph{{A very intense neutrino super beam
  experiment for leptonic CP violation discovery based on the European
  spallation source linac}},
  \href{https://doi.org/10.1016/j.nuclphysb.2014.05.016}{\emph{Nucl. Phys. B}
  {\bfseries 885} (2014) 127}
  [\href{https://arxiv.org/abs/1309.7022}{{\ttfamily 1309.7022}}].

\bibitem{DUNE:2020ypp}
{\scshape DUNE} collaboration, \emph{{Deep Underground Neutrino Experiment
  (DUNE), Far Detector Technical Design Report, Volume II: DUNE Physics}},
  \href{https://arxiv.org/abs/2002.03005}{{\ttfamily 2002.03005}}.

\bibitem{dunefluxes}
``Dune fluxes.'' https://glaucus.crc.nd.edu/DUNEFluxes/.

\bibitem{Masud:2017bcf}
M.~Masud, M.~Bishai and P.~Mehta, \emph{{Extricating New Physics Scenarios at
  DUNE with Higher Energy Beams}},
  \href{https://doi.org/10.1038/s41598-018-36790-6}{\emph{Sci. Rep.} {\bfseries
  9} (2019) 352} [\href{https://arxiv.org/abs/1704.08650}{{\ttfamily
  1704.08650}}].

\bibitem{Rout:2020cxi}
J.~Rout, S.~Roy, M.~Masud, M.~Bishai and P.~Mehta, \emph{{Impact of high energy
  beam tunes on the sensitivities to the standard unknowns at DUNE}},
  \href{https://doi.org/10.1103/PhysRevD.102.116018}{\emph{Phys. Rev. D}
  {\bfseries 102} (2020) 116018}
  [\href{https://arxiv.org/abs/2009.05061}{{\ttfamily 2009.05061}}].

\bibitem{Machado:2011jt}
P.A.N.~Machado, H.~Nunokawa and R.~Zukanovich~Funchal, \emph{{Testing for Large
  Extra Dimensions with Neutrino Oscillations}},
  \href{https://doi.org/10.1103/PhysRevD.84.013003}{\emph{Phys. Rev. D}
  {\bfseries 84} (2011) 013003}
  [\href{https://arxiv.org/abs/1101.0003}{{\ttfamily 1101.0003}}].

\bibitem{DiIura:2014csa}
A.~Di~Iura, I.~Girardi and D.~Meloni, \emph{{Probing new physics scenarios in
  accelerator and reactor neutrino experiments}},
  \href{https://doi.org/10.1088/0954-3899/42/6/065003}{\emph{J. Phys. G}
  {\bfseries 42} (2015) 065003}
  [\href{https://arxiv.org/abs/1411.5330}{{\ttfamily 1411.5330}}].

\bibitem{Berryman2016}
J.M.~Berryman, A.~de~Gouvêa, K.J.~Kelly, O.L.G.~Peres and Z.~Tabrizi,
  \emph{Large extra dimensions, the weak scale, and neutrino oscillations},
  {\emph{Physical Review D} {\bfseries 94} (2016) 033006}.

\bibitem{MINOS:2016vvv}
{\scshape MINOS} collaboration, \emph{{Constraints on Large Extra Dimensions
  from the MINOS Experiment}},
  \href{https://doi.org/10.1103/PhysRevD.94.111101}{\emph{Phys. Rev. D}
  {\bfseries 94} (2016) 111101}
  [\href{https://arxiv.org/abs/1608.06964}{{\ttfamily 1608.06964}}].

\bibitem{Evans:2017brt}
{\scshape MINOS, MINOS+} collaboration, \emph{{New results from MINOS and
  MINOS+}}, \href{https://doi.org/10.1088/1742-6596/888/1/012017}{\emph{J.
  Phys. Conf. Ser.} {\bfseries 888} (2017) 012017}.

\bibitem{Arguelles:2019xgp}
C.A.~Arg\"uelles et~al., \emph{{New opportunities at the next-generation
  neutrino experiments I: BSM neutrino physics and dark matter}},
  \href{https://doi.org/10.1088/1361-6633/ab9d12}{\emph{Rept. Prog. Phys.}
  {\bfseries 83} (2020) 124201}
  [\href{https://arxiv.org/abs/1907.08311}{{\ttfamily 1907.08311}}].

\bibitem{DUNE:2020fgq}
{\scshape DUNE} collaboration, \emph{{Prospects for beyond the Standard Model
  physics searches at the Deep Underground Neutrino Experiment}},
  \href{https://doi.org/10.1140/epjc/s10052-021-09007-w}{\emph{Eur. Phys. J. C}
  {\bfseries 81} (2021) 322}
  [\href{https://arxiv.org/abs/2008.12769}{{\ttfamily 2008.12769}}].

\bibitem{Forero:2022skg}
D.V.~Forero, C.~Giunti, C.A.~Ternes and O.~Tyagi, \emph{{Large extra dimensions
  and neutrino experiments}},
  \href{https://doi.org/10.1103/PhysRevD.106.035027}{\emph{Phys. Rev. D}
  {\bfseries 106} (2022) 035027}
  [\href{https://arxiv.org/abs/2207.02790}{{\ttfamily 2207.02790}}].

\bibitem{Khan:2022bcl}
A.N.~Khan, \emph{{Extra dimensions with light and heavy neutral leptons: an
  application to CE\ensuremath{\nu}NS}},
  \href{https://doi.org/10.1007/JHEP01(2023)052}{\emph{JHEP} {\bfseries 01}
  (2023) 052} [\href{https://arxiv.org/abs/2208.09584}{{\ttfamily
  2208.09584}}].

\bibitem{Roy:2023dyq}
S.~Roy, \emph{{Capability of the proposed long-baseline experiments to probe
  large extra dimension}},
  \href{https://doi.org/10.1103/PhysRevD.108.055015}{\emph{Phys. Rev. D}
  {\bfseries 108} (2023) 055015}
  [\href{https://arxiv.org/abs/2305.16234}{{\ttfamily 2305.16234}}].

\bibitem{Giarnetti:2024mdt}
A.~Giarnetti, S.~Marciano and D.~Meloni, \emph{{Exploring New Physics with Deep
  Underground Neutrino Experiment High-Energy Flux: The Case of Lorentz
  Invariance Violation, Large Extra Dimensions and Long-Range Forces}},
  \href{https://doi.org/10.3390/universe10090357}{\emph{Universe} {\bfseries
  10} (2024) 357} [\href{https://arxiv.org/abs/2407.17247}{{\ttfamily
  2407.17247}}].

\bibitem{MINOS:2017cae}
{\scshape MINOS+} collaboration, \emph{{Search for sterile neutrinos in MINOS
  and MINOS+ using a two-detector fit}},
  \href{https://doi.org/10.1103/PhysRevLett.122.091803}{\emph{Phys. Rev. Lett.}
  {\bfseries 122} (2019) 091803}
  [\href{https://arxiv.org/abs/1710.06488}{{\ttfamily 1710.06488}}].

\bibitem{DayaBay:2012fng}
{\scshape Daya Bay} collaboration, \emph{{Observation of electron-antineutrino
  disappearance at Daya Bay}},
  \href{https://doi.org/10.1103/PhysRevLett.108.171803}{\emph{Phys. Rev. Lett.}
  {\bfseries 108} (2012) 171803}
  [\href{https://arxiv.org/abs/1203.1669}{{\ttfamily 1203.1669}}].

\bibitem{KATRIN:2001ttj}
{\scshape KATRIN} collaboration, \emph{{KATRIN: A Next generation tritium beta
  decay experiment with sub-eV sensitivity for the electron neutrino mass.
  Letter of intent}},  \href{https://arxiv.org/abs/hep-ex/0109033}{{\ttfamily
  hep-ex/0109033}}.

\bibitem{DUNE:2021cuw}
{\scshape DUNE} collaboration, \emph{{Experiment Simulation Configurations
  Approximating DUNE TDR}},  \href{https://arxiv.org/abs/2103.04797}{{\ttfamily
  2103.04797}}.

\bibitem{Davoudiasl2002}
H.~Davoudiasl, P.~Langacker and M.~Perelstein, \emph{Constraints on large extra
  dimensions from neutrino oscillation experiments}, {\emph{Physical Review D}
  {\bfseries 65} (2002) 105015}.

\bibitem{Esmaili:2014esa}
A.~Esmaili, O.L.G.~Peres and Z.~Tabrizi, \emph{{Probing Large Extra Dimensions
  With IceCube}},
  \href{https://doi.org/10.1088/1475-7516/2014/12/002}{\emph{JCAP} {\bfseries
  12} (2014) 002} [\href{https://arxiv.org/abs/1409.3502}{{\ttfamily
  1409.3502}}].

\bibitem{Machado:2011kt}
P.A.N.~Machado, H.~Nunokawa, F.A.P.~dos Santos and R.Z.~Funchal, \emph{{Bulk
  Neutrinos as an Alternative Cause of the Gallium and Reactor Anti-neutrino
  Anomalies}}, \href{https://doi.org/10.1103/PhysRevD.85.073012}{\emph{Phys.
  Rev. D} {\bfseries 85} (2012) 073012}
  [\href{https://arxiv.org/abs/1107.2400}{{\ttfamily 1107.2400}}].

\bibitem{Basto-Gonzalez:2012nel}
V.S.~Basto-Gonzalez, A.~Esmaili and O.L.G.~Peres, \emph{{Kinematical Test of
  Large Extra Dimension in Beta Decay Experiments}},
  \href{https://doi.org/10.1016/j.physletb.2012.11.048}{\emph{Phys. Lett. B}
  {\bfseries 718} (2013) 1020}
  [\href{https://arxiv.org/abs/1205.6212}{{\ttfamily 1205.6212}}].

\bibitem{Girardi:2014gna}
I.~Girardi and D.~Meloni, \emph{{Constraining new physics scenarios in neutrino
  oscillations from Daya Bay data}},
  \href{https://doi.org/10.1103/PhysRevD.90.073011}{\emph{Phys. Rev. D}
  {\bfseries 90} (2014) 073011}
  [\href{https://arxiv.org/abs/1403.5507}{{\ttfamily 1403.5507}}].

\bibitem{Rodejohann:2014eka}
W.~Rodejohann and H.~Zhang, \emph{{Signatures of Extra Dimensional Sterile
  Neutrinos}},
  \href{https://doi.org/10.1016/j.physletb.2014.08.035}{\emph{Phys. Lett. B}
  {\bfseries 737} (2014) 81} [\href{https://arxiv.org/abs/1407.2739}{{\ttfamily
  1407.2739}}].

\bibitem{Carena:2017qhd}
M.~Carena, Y.-Y.~Li, C.S.~Machado, P.A.N.~Machado and C.E.M.~Wagner,
  \emph{{Neutrinos in Large Extra Dimensions and Short-Baseline $\nu_e$
  Appearance}}, \href{https://doi.org/10.1103/PhysRevD.96.095014}{\emph{Phys.
  Rev. D} {\bfseries 96} (2017) 095014}
  [\href{https://arxiv.org/abs/1708.09548}{{\ttfamily 1708.09548}}].

\bibitem{Stenico:2018jpl}
G.V.~Stenico, D.V.~Forero and O.L.G.~Peres, \emph{{A Short Travel for Neutrinos
  in Large Extra Dimensions}},
  \href{https://doi.org/10.1007/JHEP11(2018)155}{\emph{JHEP} {\bfseries 11}
  (2018) 155} [\href{https://arxiv.org/abs/1808.05450}{{\ttfamily
  1808.05450}}].

\bibitem{Basto-Gonzalez:2021aus}
V.S.~Basto-Gonzalez, D.V.~Forero, C.~Giunti, A.A.~Quiroga and C.A.~Ternes,
  \emph{{Short-baseline oscillation scenarios at JUNO and TAO}},
  \href{https://doi.org/10.1103/PhysRevD.105.075023}{\emph{Phys. Rev. D}
  {\bfseries 105} (2022) 075023}
  [\href{https://arxiv.org/abs/2112.00379}{{\ttfamily 2112.00379}}].

\bibitem{Mohapatra:2000wn}
R.N.~Mohapatra and A.~Perez-Lorenzana, \emph{{Three flavor neutrino
  oscillations in models with large extra dimensions}},
  \href{https://doi.org/10.1016/S0550-3213(00)00634-9}{\emph{Nucl. Phys. B}
  {\bfseries 593} (2001) 451}
  [\href{https://arxiv.org/abs/hep-ph/0006278}{{\ttfamily hep-ph/0006278}}].

\bibitem{Huber:2004ka}
P.~Huber, M.~Lindner and W.~Winter, \emph{{Simulation of long-baseline neutrino
  oscillation experiments with GLoBES (General Long Baseline Experiment
  Simulator)}}, \href{https://doi.org/10.1016/j.cpc.2005.01.003}{\emph{Comput.
  Phys. Commun.} {\bfseries 167} (2005) 195}
  [\href{https://arxiv.org/abs/hep-ph/0407333}{{\ttfamily hep-ph/0407333}}].

\bibitem{Huber:2007ji}
P.~Huber, J.~Kopp, M.~Lindner, M.~Rolinec and W.~Winter, \emph{{New features in
  the simulation of neutrino oscillation experiments with GLoBES 3.0: General
  Long Baseline Experiment Simulator}},
  \href{https://doi.org/10.1016/j.cpc.2007.05.004}{\emph{Comput. Phys. Commun.}
  {\bfseries 177} (2007) 432}
  [\href{https://arxiv.org/abs/hep-ph/0701187}{{\ttfamily hep-ph/0701187}}].

\bibitem{Agostinelli:2002hh}
{\scshape GEANT4} collaboration, \emph{{GEANT4: A Simulation toolkit}},
  \href{https://doi.org/10.1016/S0168-9002(03)01368-8}{\emph{Nucl. Instrum.
  Meth.} {\bfseries A506} (2003) 250}.

\bibitem{Allison:2006ve}
J.~Allison et~al., \emph{{Geant4 developments and applications}},
  \href{https://doi.org/10.1109/TNS.2006.869826}{\emph{IEEE Trans. Nucl. Sci.}
  {\bfseries 53} (2006) 270}.

\bibitem{deGouvea:2019ozk}
A.~De~Gouv\^ea, K.J.~Kelly, G.~Stenico and P.~Pasquini, \emph{{Physics with
  Beam Tau-Neutrino Appearance at DUNE}},
  \href{https://doi.org/10.1103/PhysRevD.100.016004}{\emph{Phys. Rev. D}
  {\bfseries 100} (2019) 016004}
  [\href{https://arxiv.org/abs/1904.07265}{{\ttfamily 1904.07265}}].

\bibitem{Ghoshal:2019pab}
A.~Ghoshal, A.~Giarnetti and D.~Meloni, \emph{{On the role of the $\nu_{?}$
  appearance in DUNE in constraining standard neutrino physics and beyond}},
  \href{https://doi.org/10.1007/JHEP12(2019)126}{\emph{JHEP} {\bfseries 12}
  (2019) 126} [\href{https://arxiv.org/abs/1906.06212}{{\ttfamily
  1906.06212}}].

\bibitem{Machado:2020yxl}
P.~Machado, H.~Schulz and J.~Turner, \emph{{Tau neutrinos at DUNE: New
  strategies, new opportunities}},
  \href{https://doi.org/10.1103/PhysRevD.102.053010}{\emph{Phys. Rev. D}
  {\bfseries 102} (2020) 053010}
  [\href{https://arxiv.org/abs/2007.00015}{{\ttfamily 2007.00015}}].

\bibitem{DUNE:2016ymp}
{\scshape DUNE} collaboration, \emph{{Experiment Simulation Configurations Used
  in DUNE CDR}},  \href{https://arxiv.org/abs/1606.09550}{{\ttfamily
  1606.09550}}.

\bibitem{Huber:2002mx}
P.~Huber, M.~Lindner and W.~Winter, \emph{{Superbeams versus neutrino
  factories}}, \href{https://doi.org/10.1016/S0550-3213(02)00825-8}{\emph{Nucl.
  Phys.} {\bfseries B645} (2002) 3}
  [\href{https://arxiv.org/abs/hep-ph/0204352}{{\ttfamily hep-ph/0204352}}].

\bibitem{Fogli:2002pt}
G.L.~Fogli, E.~Lisi, A.~Marrone, D.~Montanino and A.~Palazzo, \emph{{Getting
  the most from the statistical analysis of solar neutrino oscillations}},
  \href{https://doi.org/10.1103/PhysRevD.66.053010}{\emph{Phys. Rev.}
  {\bfseries D66} (2002) 053010}
  [\href{https://arxiv.org/abs/hep-ph/0206162}{{\ttfamily hep-ph/0206162}}].

\bibitem{GonzalezGarcia:2004wg}
M.~Gonzalez-Garcia and M.~Maltoni, \emph{{Atmospheric neutrino oscillations and
  new physics}},
  \href{https://doi.org/10.1103/PhysRevD.70.033010}{\emph{Phys.Rev.} {\bfseries
  D70} (2004) 033010} [\href{https://arxiv.org/abs/hep-ph/0404085}{{\ttfamily
  hep-ph/0404085}}].

\bibitem{Gandhi:2007td}
R.~Gandhi, P.~Ghoshal, S.~Goswami, P.~Mehta, S.U.~Sankar and S.~Shalgar,
  \emph{{Mass Hierarchy Determination via future Atmospheric Neutrino
  Detectors}}, \href{https://doi.org/10.1103/PhysRevD.76.073012}{\emph{Phys.
  Rev.} {\bfseries D76} (2007) 073012}
  [\href{https://arxiv.org/abs/0707.1723}{{\ttfamily 0707.1723}}].

\bibitem{Qian:2012zn}
X.~Qian, A.~Tan, W.~Wang, J.J.~Ling, R.D.~McKeown and C.~Zhang,
  \emph{{Statistical Evaluation of Experimental Determinations of Neutrino Mass
  Hierarchy}}, \href{https://doi.org/10.1103/PhysRevD.86.113011}{\emph{Phys.
  Rev.} {\bfseries D86} (2012) 113011}
  [\href{https://arxiv.org/abs/1210.3651}{{\ttfamily 1210.3651}}].

\bibitem{GALLEX:1997lja}
{\scshape GALLEX} collaboration, \emph{{Final results of the Cr-51 neutrino
  source experiments in GALLEX}},
  \href{https://doi.org/10.1016/S0370-2693(97)01562-1}{\emph{Phys. Lett. B}
  {\bfseries 420} (1998) 114}.

\bibitem{SAGE:1998fvr}
{\scshape SAGE} collaboration, \emph{{Measurement of the response of the
  Russian-American gallium experiment to neutrinos from a Cr-51 source}},
  \href{https://doi.org/10.1103/PhysRevC.59.2246}{\emph{Phys. Rev. C}
  {\bfseries 59} (1999) 2246}
  [\href{https://arxiv.org/abs/hep-ph/9803418}{{\ttfamily hep-ph/9803418}}].

\bibitem{Barinov:2021asz}
V.V.~Barinov et~al., \emph{{Results from the Baksan Experiment on Sterile
  Transitions (BEST)}},
  \href{https://doi.org/10.1103/PhysRevLett.128.232501}{\emph{Phys. Rev. Lett.}
  {\bfseries 128} (2022) 232501}
  [\href{https://arxiv.org/abs/2109.11482}{{\ttfamily 2109.11482}}].

\bibitem{Long:1998dk}
J.C.~Long, H.W.~Chan and J.C.~Price, \emph{{Experimental status of
  gravitational strength forces in the subcentimeter regime}},
  \href{https://doi.org/10.1016/S0550-3213(98)00711-1}{\emph{Nucl. Phys. B}
  {\bfseries 539} (1999) 23}
  [\href{https://arxiv.org/abs/hep-ph/9805217}{{\ttfamily hep-ph/9805217}}].

\bibitem{Krause:1999ry}
D.E.~Krause and E.~Fischbach, \emph{{Searching for extra dimensions and new
  string inspired forces in the Casimir regime}},
  \href{https://doi.org/10.1007/3-540-40988-2_14}{\emph{Lect. Notes Phys.}
  {\bfseries 562} (2001) 292}
  [\href{https://arxiv.org/abs/hep-ph/9912276}{{\ttfamily hep-ph/9912276}}].

\bibitem{Fischbach:2001ry}
E.~Fischbach, D.E.~Krause, V.M.~Mostepanenko and M.~Novello, \emph{{New
  constraints on ultrashort ranged Yukawa interactions from atomic force
  microscopy}}, \href{https://doi.org/10.1103/PhysRevD.64.075010}{\emph{Phys.
  Rev. D} {\bfseries 64} (2001) 075010}
  [\href{https://arxiv.org/abs/hep-ph/0106331}{{\ttfamily hep-ph/0106331}}].

\bibitem{Adelberger:2002ic}
{\scshape EOT-WASH Group} collaboration, \emph{{Sub-millimeter tests of the
  gravitational inverse square law}},  in \emph{{2nd Meeting on CPT and Lorentz
  Symmetry}}, pp.~9--15, 2002,
  \href{https://doi.org/10.1142/9789812778123_0002}{DOI}
  [\href{https://arxiv.org/abs/hep-ex/0202008}{{\ttfamily hep-ex/0202008}}].

\bibitem{Decca:2003td}
R.S.~Decca, E.~Fischbach, G.L.~Klimchitskaya, D.E.~Krause, D.L.~Lopez and
  V.M.~Mostepanenko, \emph{{Improved tests of extra dimensional physics and
  thermal quantum field theory from new Casimir force measurements}},
  \href{https://doi.org/10.1103/PhysRevD.68.116003}{\emph{Phys. Rev. D}
  {\bfseries 68} (2003) 116003}
  [\href{https://arxiv.org/abs/hep-ph/0310157}{{\ttfamily hep-ph/0310157}}].

\bibitem{Rizzo:1998fm}
T.G.~Rizzo, \emph{{More and more indirect signals for extra dimensions at more
  and more colliders}},
  \href{https://doi.org/10.1103/PhysRevD.59.115010}{\emph{Phys. Rev. D}
  {\bfseries 59} (1999) 115010}
  [\href{https://arxiv.org/abs/hep-ph/9901209}{{\ttfamily hep-ph/9901209}}].

\bibitem{Hewett:1998sn}
J.L.~Hewett, \emph{{Indirect collider signals for extra dimensions}},
  \href{https://doi.org/10.1103/PhysRevLett.82.4765}{\emph{Phys. Rev. Lett.}
  {\bfseries 82} (1999) 4765}
  [\href{https://arxiv.org/abs/hep-ph/9811356}{{\ttfamily hep-ph/9811356}}].

\bibitem{D0:2000cve}
{\scshape D0} collaboration, \emph{{Search for large extra dimensions in
  dielectron and diphoton production}},
  \href{https://doi.org/10.1103/PhysRevLett.86.1156}{\emph{Phys. Rev. Lett.}
  {\bfseries 86} (2001) 1156}
  [\href{https://arxiv.org/abs/hep-ex/0008065}{{\ttfamily hep-ex/0008065}}].

\bibitem{DELPHI:2000ztm}
{\scshape DELPHI} collaboration, \emph{{Measurement and interpretation of
  fermion-pair production at LEP energies of 183-GeV and 189-GeV}},
  \href{https://doi.org/10.1016/S0370-2693(00)00675-4}{\emph{Phys. Lett. B}
  {\bfseries 485} (2000) 45}
  [\href{https://arxiv.org/abs/hep-ex/0103025}{{\ttfamily hep-ex/0103025}}].

\bibitem{DELPHI:2008uka}
{\scshape DELPHI} collaboration, \emph{{Search for one large extra dimension
  with the DELPHI detector at LEP}},
  \href{https://doi.org/10.1140/epjc/s10052-009-0874-9}{\emph{Eur. Phys. J. C}
  {\bfseries 60} (2009) 17} [\href{https://arxiv.org/abs/0901.4486}{{\ttfamily
  0901.4486}}].

\bibitem{Cullen:1999hc}
S.~Cullen and M.~Perelstein, \emph{{SN1987A constraints on large compact
  dimensions}}, \href{https://doi.org/10.1103/PhysRevLett.83.268}{\emph{Phys.
  Rev. Lett.} {\bfseries 83} (1999) 268}
  [\href{https://arxiv.org/abs/hep-ph/9903422}{{\ttfamily hep-ph/9903422}}].

\bibitem{Barger:1999jf}
V.D.~Barger, T.~Han, C.~Kao and R.-J.~Zhang, \emph{{Astrophysical constraints
  on large extra dimensions}},
  \href{https://doi.org/10.1016/S0370-2693(99)00795-9}{\emph{Phys. Lett. B}
  {\bfseries 461} (1999) 34}
  [\href{https://arxiv.org/abs/hep-ph/9905474}{{\ttfamily hep-ph/9905474}}].

\bibitem{Hanhart:2000er}
C.~Hanhart, D.R.~Phillips, S.~Reddy and M.J.~Savage, \emph{{Extra dimensions,
  SN1987a, and nucleon-nucleon scattering data}},
  \href{https://doi.org/10.1016/S0550-3213(00)00667-2}{\emph{Nucl. Phys. B}
  {\bfseries 595} (2001) 335}
  [\href{https://arxiv.org/abs/nucl-th/0007016}{{\ttfamily nucl-th/0007016}}].

\bibitem{Hannestad:2001jv}
S.~Hannestad and G.~Raffelt, \emph{{New supernova limit on large extra
  dimensions}},
  \href{https://doi.org/10.1103/PhysRevLett.87.051301}{\emph{Phys. Rev. Lett.}
  {\bfseries 87} (2001) 051301}
  [\href{https://arxiv.org/abs/hep-ph/0103201}{{\ttfamily hep-ph/0103201}}].

\bibitem{Hannestad:2003yd}
S.~Hannestad and G.G.~Raffelt, \emph{{Supernova and neutron star limits on
  large extra dimensions reexamined}},
  \href{https://doi.org/10.1103/PhysRevD.69.029901}{\emph{Phys. Rev. D}
  {\bfseries 67} (2003) 125008}
  [\href{https://arxiv.org/abs/hep-ph/0304029}{{\ttfamily hep-ph/0304029}}].

\bibitem{Mohapatra:2003ah}
R.N.~Mohapatra, S.~Nussinov and A.~Perez-Lorenzana, \emph{{Large extra
  dimensions and decaying $K K$ recurrences}},
  \href{https://doi.org/10.1103/PhysRevD.68.116001}{\emph{Phys. Rev. D}
  {\bfseries 68} (2003) 116001}
  [\href{https://arxiv.org/abs/hep-ph/0308051}{{\ttfamily hep-ph/0308051}}].

\bibitem{Feng:2003nr}
J.L.~Feng, A.~Rajaraman and F.~Takayama, \emph{{Graviton cosmology in universal
  extra dimensions}},
  \href{https://doi.org/10.1103/PhysRevD.68.085018}{\emph{Phys. Rev. D}
  {\bfseries 68} (2003) 085018}
  [\href{https://arxiv.org/abs/hep-ph/0307375}{{\ttfamily hep-ph/0307375}}].

\bibitem{Cacciapaglia:2003dx}
G.~Cacciapaglia, M.~Cirelli and A.~Romanino, \emph{{Signatures of supernova
  neutrino oscillations into extra dimensions}},
  \href{https://doi.org/10.1103/PhysRevD.68.033013}{\emph{Phys. Rev. D}
  {\bfseries 68} (2003) 033013}
  [\href{https://arxiv.org/abs/hep-ph/0302246}{{\ttfamily hep-ph/0302246}}].

\bibitem{Hall:1999mk}
L.J.~Hall and D.~Tucker-Smith, \emph{{Cosmological constraints on theories with
  large extra dimensions}},
  \href{https://doi.org/10.1103/PhysRevD.60.085008}{\emph{Phys. Rev. D}
  {\bfseries 60} (1999) 085008}
  [\href{https://arxiv.org/abs/hep-ph/9904267}{{\ttfamily hep-ph/9904267}}].

\bibitem{Hannestad:2001nq}
S.~Hannestad, \emph{{Strong constraint on large extra dimensions from
  cosmology}}, \href{https://doi.org/10.1103/PhysRevD.64.023515}{\emph{Phys.
  Rev. D} {\bfseries 64} (2001) 023515}
  [\href{https://arxiv.org/abs/hep-ph/0102290}{{\ttfamily hep-ph/0102290}}].

\bibitem{Fairbairn:2001ct}
M.~Fairbairn, \emph{{Cosmological constraints on large extra dimensions}},
  \href{https://doi.org/10.1016/S0370-2693(01)00501-9}{\emph{Phys. Lett. B}
  {\bfseries 508} (2001) 335}
  [\href{https://arxiv.org/abs/hep-ph/0101131}{{\ttfamily hep-ph/0101131}}].

\bibitem{Fairbairn:2001ab}
M.~Fairbairn and L.M.~Griffiths, \emph{{Large extra dimensions, the galaxy
  power spectrum and the end of inflation}},
  \href{https://doi.org/10.1088/1126-6708/2002/02/024}{\emph{JHEP} {\bfseries
  02} (2002) 024} [\href{https://arxiv.org/abs/hep-ph/0111435}{{\ttfamily
  hep-ph/0111435}}].

\bibitem{Goh:2001uc}
H.S.~Goh and R.N.~Mohapatra, \emph{{Big bang nucleosynthesis constraints on
  bulk neutrinos}},
  \href{https://doi.org/10.1103/PhysRevD.65.085018}{\emph{Phys. Rev. D}
  {\bfseries 65} (2002) 085018}
  [\href{https://arxiv.org/abs/hep-ph/0110161}{{\ttfamily hep-ph/0110161}}].

\bibitem{ParticleDataGroup:2022pth}
{\scshape Particle Data Group} collaboration, \emph{{Review of Particle
  Physics}}, \href{https://doi.org/10.1093/ptep/ptac097}{\emph{PTEP} {\bfseries
  2022} (2022) 083C01}.

\bibitem{Sanderson2016}
C.~Sanderson and R.~Curtin, \emph{Armadillo: a template-based c++ library for
  linear algebra}, \href{https://doi.org/10.21105/joss.00026}{\emph{Journal of
  Open Source Software} {\bfseries 1} (2016) 26}.

\bibitem{mca24030070}
C.~Sanderson and R.~Curtin, \emph{Practical sparse matrices in c++ with hybrid
  storage and template-based expression optimisation},
  \href{https://doi.org/10.3390/mca24030070}{\emph{Mathematical and
  Computational Applications} {\bfseries 24} (2019) }.

\end{thebibliography}\endgroup
\end{document}